\documentclass[11pt]{iopart}

\usepackage{graphicx}
\usepackage{amssymb}

\usepackage{float}
\usepackage{tabularx}
\usepackage{microtype}
\usepackage{hyperref}

\usepackage{enumitem}
\usepackage{xcolor}
\usepackage{cite}

\usepackage{etoolbox}

\makeatletter
\renewcommand\subsubsection{\@startsection{subsubsection}{3}{\z@}%
                                    {-3.25ex\@plus -1ex \@minus -.2ex}%
                                    {1.5ex \@plus .2ex}%
                                    {\normalfont\normalsize\itshape\raggedright}}
\patchcmd{\subsubsection}{.5em}{\newline}{}{} 
\makeatother

\begin{document}



\title[]{Investigations of atomic and molecular processes of NBI-heated discharges in the MAST Upgrade Super-X divertor with implications for reactors}



\author{K. Verhaegh$^{1}$, J.R. Harrison$^1$, B. Lipschultz$^2$, N. Lonigro$^{2,1}$, S. Kobussen$^3$, D. Moulton$^1$, N. Osborne$^{4,1}$, P. Ryan$^1$, C. Theiler$^5$, T. Wijkamp$^{3,6}$, D. Brida$^7$, G. Derks$^6$, R. Doyle$^8$, F. Federici$^{9,1}$,A. Hakola$^{10}$, S. Henderson$^1$, B. Kool$^6$, S. Newton$^1$, R. Osawa$^1$, X. Pope$^{2,1}$, H. Reimerdes$^5$, N. Vianello$^{10}$, M. Wischmeier$^8$ and the EUROfusion Tokamak Exploitation Team$^{*}$ and the MAST-U Team$^{**}$}
\address{$^1$ United Kingdom Atomic Energy Authority, Culham, United Kingdom} 
\address{$^2$ York Plasma Institute, University of York, United Kingdom}
\address{$^3$ Eindhoven University of Technology, Eindhoven, The Netherlands}
\address{$^4$ University of Liverpool, Liverpool, United Kingdom}
\address{$^5$ Swiss Plasma Centre, \'{E}cole Polytechnique F\'{e}d\'{e}rale de Lausanne, Lausanne, Switzerland}
\address{$^6$ Dutch Institute for Fundamental Energy Research DIFFER, Eindhoven, The Netherlands}
\address{$^7$ Max Planck Institute, Garching, Germany, The Netherlands}
\address{$^8$ Dublin City University, Dublin, Ireland}
\address{$^9$ Oak Ridge National Laboratory, Oak Ridge, United States}
\address{$^{10}$ VTT, Espoo, Finland}
\address{$^{11}$ Consorzio RFX, Padova, Italy}
\address{$^{*}$ See the author list of “Progress on an exhaust solution for a reactor using EUROfusion multi-machines capabilities” by E. Joffrin et al. to be published in Nuclear Fusion Special Issue: Overview and Summary Papers from the 29th Fusion Energy Conference (London, UK, 16-21 October 2023).}
\address{$^{**}$ See the author list of “Overview of physics results from MAST Upgrade towards
core-pedestal-exhaust integration” by J.R. Harrison et al. to be published in Nuclear Fusion Special Issue: Overview and Summary Papers from the 29th Fusion Energy Conference (London, UK, 16-21 October 2023).}



\ead{kevin.verhaegh@ukaea.uk}

\begin{abstract}
This experimental study presents an in-depth investigation of the performance of the MAST-U Super-X divertor during NBI-heated operation (up to 2.5 MW) focussing on volumetric ion sources and sinks as well as power losses during detachment.

The particle balance and power loss analysis revealed the crucial role of Molecular Activated Recombination and Dissociation (MAR and MAD) ion sinks in divertor particle and power balance, which remain pronounced in the change from ohmic to higher power (NBI heated) L-mode conditions. The importance of MAR and MAD remains with double the absorbed NBI heating. MAD results in significant power dissipation (up to $\sim 20 \%$ of $P_{SOL}$), mostly in the cold ($T_e < 5$ eV) detached region. Theoretical and experimental evidence is found for the potential contribution of $D^-$ to MAR and MAD, which warrants further study. 

These results suggest that MAR and MAD can be relevant in higher power conditions than the ohmic conditions studied previously. Post-processing reactor-scale simulations shows that MAR and MAD can play a significant role in divertor physics and synthetic diagnostic signals of reactor-scale devices, which are currently underestimated in exhaust simulations. This raises implications for the accuracy of reactor-scale divertor simulations of particularly tightly baffled (alternative) divertor configurations.
\end{abstract}


\section{Introduction}
\label{ch:introduction}

Efficient power exhaust is a paramount challenge in the realisation of fusion energy, particularly in reactor designs such as DEMO \cite{Wenninger2014} and STEP \cite{Wilson2020}. In these designs, heat fluxes can escalate beyond engineering limits due to surface recombination alone, presenting a formidable obstacle to the practical application of fusion energy \cite{Asakura2013}. Plasma detachment, achieved through various plasma-neutral interactions, stands out as a promising solution to mitigate plasma target heat and ion fluxes \cite{Verhaegh2021b} through power, particle (e.g. ion), and momentum removal. The reduction in ion target flux ($I_t$) is a crucial element of plasma detachment and can be obtained through ion sinks ($I_r$), which neutralise ions before they reach the target, and/or by reducing/limiting the total ion source (divertor ionisation $I_i$ plus an inflow of ions from upstream if present ($I_u$)) through power limitation; see equation \ref{eq:PartBal} \cite{Verhaegh2019, Lipschultz1999, Krasheninnikov2017, Verhaegh2017, Verhaegh2021b}.

\begin{equation}
I_t = I_i - I_r + I_u
\label{eq:PartBal}
\end{equation}

However, uncertainties persist as to whether sufficiently low target heat fluxes with a sufficient core performance can be achieved in tokamak reactor designs. To confront this challenge, researchers are exploring alternative divertor configurations, leveraging magnetic shaping \cite{Theiler2017,Reimerdes2017,Verhaegh2023b,Ryutov2015,Kotschenreuther2013,Soukhanovskii2009} and robust baffling \cite{Reimerdes2021,Verhaegh2023b,Sang2017} to enhance power exhaust efficiency.

The Super-X divertor configuration, as realised in MAST-Upgrade, integrates strong divertor baffling with high 'total flux expansion' ($f_R = B_{tot, xpt}/B_{tot, t} > 2.5$) \cite{Moulton2023, Harrison2023,Verhaegh2023b,Verhaegh2023d}, defined as the ratio between the total magnetic field at the X-point and target \footnote{This follows the definition from \cite{Lipschultz2016,Theiler2017}, also referred to as the 'parallel area expansion' \cite{Moulton2017}, not to be confused with poloidal flux expansion $f_x = \frac{B_{\theta, omp} B_{\phi, t}}{B_{\theta, t} B_{\phi, omp}}$ \cite{Theiler2017}, defined 'as the ratio of the perpendicular flux surface spacing at the target and upstream' \cite{Theiler2017}, where $B_{\theta, \phi}$ are the poloidal and toroidal magnetic field and omp, t corresponds to outer midplane and target locations, respectively. Figure 1b,c in \cite{Theiler2017} illustrates the differences between total and poloidal flux expansion.}. Ohmic experiments ($P_{SOL} \approx 0.5-0.6$ MW) show this can reduce target heat fluxes by more than tenfold and facilitate access to plasma detachment \cite{Moulton2023,Harrison2023,Verhaegh2023b,Verhaegh2023}. In these experiments, Molecular Activated Recombination (MAR) \cite{Verhaegh2021, Verhaegh2021b, Verhaegh2023} emerged as the dominant mechanism to kerb ion target fluxes and facilitate detachment in the divertor \cite{Verhaegh2023, Verhaegh2023b, Wijkamp2023}. Furthermore, Molecular Activated Dissociation (MAD) has been identified as a significant contributor to electron cooling, accounting for a substantial fraction of $P_{SOL}$ ($>10 \%$).

 This paper presents a comprehensive investigation into the detachment mechanisms as well as power and particle balance of the MAST-U \cite{Harrison2023a} Super-X divertor under higher-power, neutral beam-heated (up to 2.5 MW), L-mode conditions. This is achieved using hydrogen emission line-of-sight spectroscopy \cite{Verhaegh2023,Verhaegh2023b} and multi-wavelength imaging \cite{Wijkamp2023}, in combination with state-of-the-art spectroscopic analysis \cite{Verhaegh2019a,Verhaegh2023,Verhaegh2023b}. Our findings confirm the importance of MAR and MAD in these higher power scenarios, mirroring the results in ohmically heated conditions \cite{Verhaegh2023,Verhaegh2023b}. Evidence is found for the potential contribution of $D^-$ to MAR and MAD, a previously overlooked factor.

Our analysis, through post-processing reactor-relevant exhaust simulations, shows that MAR and MAD can be significant in reactor conditions. Plasma-molecular chemistry, involving $D_2^+$ and $D_2^- \rightarrow D + D^-$, is generally underestimated in exhaust simulations under low temperature conditions ($<2$ eV) as $D^-$ is neglected and the molecular charge exchange rates ($D_2 + D^+ \rightarrow D_2^+ + D$) currently used \cite{Reiter2005,Greenland2001,Holliday1971}, are underestimated at high vibrational levels ($\nu \geq 4$) by up to orders of magnitude for $T<3$ eV \cite{Verhaegh2021b,Verhaegh2023a,Ichihara2000}. Our work shows that these underpredictions can result in uncertainties in reactor predictions for divertor heat loads; divertor particle (ion and neutral) balance; as well as hydrogen emission synthetic diagnostic predictions. Improvements are therefore essential to reduce uncertainties in extrapolating our current understanding to reactor-class devices, particularly those featuring tightly baffled alternative divertor designs.

\section{Experimental setup}
\label{ch:MASTU_overview}

\begin{table}[]
\begin{tabular}{lllll}
Discharge                       & 46860        & 46705      & 47958                                                                 & 48330                                                                      \\
$P_{NBI}^{on-axis}$ (MW)        & 0            & 0-0.9      & 0                                                                     & 0                                                                          \\
$P_{NBI}^{off-axis}$ (MW)       & 1.7          & 1.7-0      & 1.6                                                                   & 1.6                                                                        \\
$P_{NBI}^{absorbed}$ (MW)       & 0.6-1.0      & 0.8-1.7    & 0.5-1.0                                                               & 0.5-1.0                                                                    \\
$P_{SOL}$ (MW)                  & 1.0-1.3      & 1.2-$<2.1$ & 0.9-1.3                                                               & 0.9-1.3                                                                    \\
$P_{rad}^{core}$                & 0.2-0.35     & 0.2        & 0.1-0.35                                                              & 0.2-0.35                                                                   \\
$f_{GW}$                        & 0.27-0.47    & 0.3        & 0.22-0.45                                                             & 0.2-0.42                                                                   \\
Core $T_e$ (keV)                & 1-0.6        & 1-1.15     & 1.1-0.6                                                               & 1.1-0.7                                                                    \\
Fuelling location               & LFS-V        & LFS-V      & LFS-V                                                                 & LFS-V                                                                      \\
Fuelling flux ($10^{21}$ /s) & 0.4 - 7.3    & $>0^*$     & 1.1-7.3                                                               & 1.1-7.3                                                                    \\
Description                     & Density ramp & Power scan & \begin{tabular}[c]{@{}l@{}}Density ramp \\ (IRVB repeat)\end{tabular} & \begin{tabular}[c]{@{}l@{}}Density ramp \\ (Xpt imag. repeat)\end{tabular}
\end{tabular}
\caption{Table with relevant discharge parameter ranges for the discharges used in this work. $^*$ implies that the gas injected was below the calibrated range (although, from imaging, it was clear that gas was injected).}
\label{tab:ShotParam}
\end{table}

This study focusses on beam-heated L-mode discharges in the Super-X configuration. To prevent the plasma from transitioning into H-mode, fuelling from the low-field side was employed. The choice of beam-heated L-mode operation served several key purposes:

\begin{itemize}
    \item Enhanced Diagnosability: Beam-heated L-mode conditions ensured optimal diagnosability. Compared to H-mode, this choice avoids complications with insufficient acquisition rates of certain diagnostics to obtain inter-ELM measurements.
    \item Control over Core/Upstream Density: This mode of operation allowed precise control over the core/upstream density, which enables investigating the progression of detachment onset to more deeply detached conditions
    \item Stable Divertor Geometry: Maintaining beam-heated L-mode conditions facilitated the stability and consistency of the divertor geometry, improving diagnosability, and facilitating code validation studies.
\end{itemize}

It should be noted that, as demonstrated in previous research \cite{Verhaegh2023b,Verhaegh2023}, divertor detachment behaviour is not expected to undergo significant changes when transitioning from L-mode to H-mode operation, provided that the power entering the scrape-off layer ($P_{SOL}$) and the upstream density remain at similar levels \cite{Verhaegh2023b}. The ability to control the core and upstream density during beam-heated L-mode conditions also allowed us to explore conditions in which the divertor remains less deeply detached than in higher-power H-mode operation, as detailed in \cite{Harrison2023}. However, studying the Super-X divertor in H-mode, as well as under higher power, conditions is an important future focus \cite{Harrison2023}. H-mode operation features transient Edge Localised Modes (ELMs), narrower heat flux widths, and enables higher $P_{SOL}$ operation than possible in L-mode when additional heating power is available. Cryopumping was not available for these discharges.

\begin{figure}[htbp]
\centering
\includegraphics[width=0.7\textwidth]{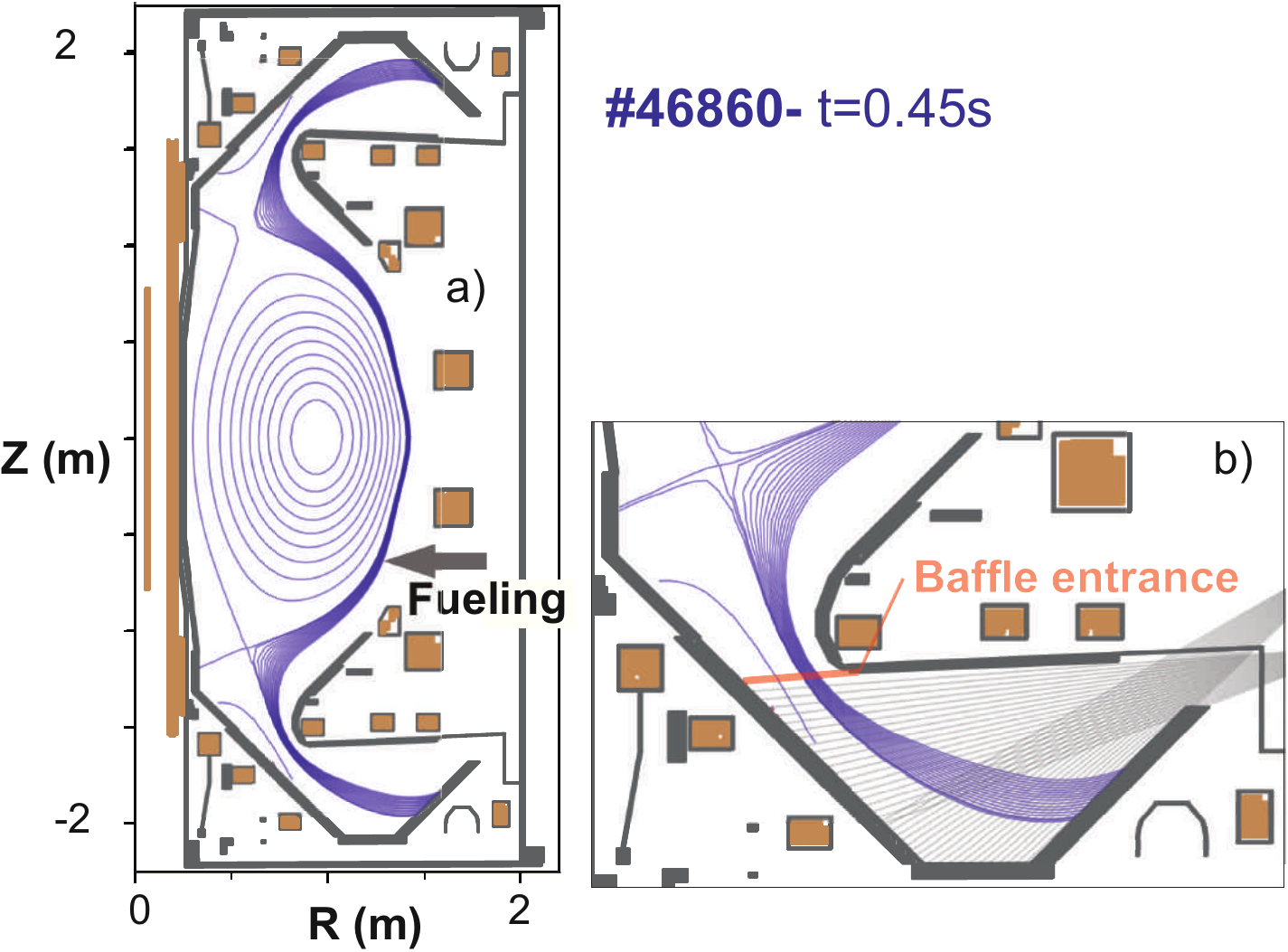}
\caption{Overview of the magnetic geometry for the Super-X Divertor. Low field side fuelling is applied to maintain L-mode under beam-heated conditions (a). Zoom-in of the lower divertor chamber highlighting the Super-X divertor geometry, together with the baffle entrance position and the lines of sight of the divertor spectrometer \cite{Verhaegh2023,Verhaegh2023b} (b).}
\label{fig:GeomOverview}
\end{figure}

The magnetic geometry employed in this work, as illustrated in figure \ref{fig:GeomOverview}, has been improved, compared to initial studies \cite{Harrison2023,Moulton2023}, to optimise it specifically for divertor physics studies. The elimination of secondary (unwanted) X-points in the divertor chamber has reduced the interaction of the far-Scrape-Off Layer (SOL) with the underside of the baffle. This ensures that the last flux tube entering the baffle entrance reaches the edge of the target tile, enhancing the diagnosability of the divertor and facilitating future interpretive SOLPS-ITER simulations \cite{Moulton2023}. Feedback control on the inner gap has been employed to increase it to 4 cm (figure \ref{fig:GeomOverview}a). Coupled with the absence of fuelling from the high field side, this adjustment has substantially reduced radiative power losses from the high field side and prevented an upstream pressure collapse in SOLPS-ITER simulations \cite{Moulton2023}.

In table \ref{tab:ShotParam}, we provide a summary of the relevant discharge parameters for the discharges analysed in this study. The main discharge that will be used throughout the paper is density ramp \# 46860, which is supplemented with data from repeat discharges for imaging bolometry coverage (\# 47958, used in figures \ref{fig:D2Fulcher_XPT} and \ref{fig:PowerLossSXD}) and X-point imaging coverage (\# 48330, only used in figure \ref{fig:D2Fulcher_XPT}). Data from \# 46705 are used to discuss the impact of doubing the absorbed beam power (figure \ref{fig:PowerTransientSXD}). 

\section{Results of the beam-heated Super-X divertor}
\label{ch:PartPowerBal}

In this section, we present the key findings of our experimental study on the MAST-U Super-X divertor, focussing on an analysis of the evolution of ion sources and sinks (section \ref{ch:IonProf}), as well as particle and power balance (section \ref{ch:PowerPart}), as the core density (represented relative to the Greenwald density limit \cite{Greenwald1988} - Greenwald fraction ($f_{GW}$)) is gradually increased to obtain deeper detachment. Afterwards, the impact of additional heating on the detached state is studied (section \ref{ch:PNBI}).

This analysis has been enabled through a novel spectroscopic technique \cite{Verhaegh2019a,Verhaegh2021,Verhaegh2023b} that uses a Bayesian approach to quantitatively infer the contribution to the hydrogen Balmer line emission due to electron impact excitation \cite{Verhaegh2019a}, electron ion recombination \cite{Verhaegh2019a}, and excited neutral atoms arising from plasma-molecular interactions \cite{Verhaegh2021,Greenhouse2024} using emission coefficients from ADAS (atomic interactions) \cite{OMullane} and Yacora(-on the Web) (molecular interactions) \cite{Wunderlich2020,Wuenderlich2016}. These hydrogen Balmer line emission contributions are post-processed to obtain quantitative information on atomic \cite{Verhaegh2019a} and molecular \cite{Verhaegh2021} reactions and hydrogenic radiative power losses. Monte Carlo sampling is applied to the outcome of the Bayesian analysis for uncertainty propagation, and the uncertainties shown correspond to 68 \% confidence intervals. More details are provided in \cite{Verhaegh2019a,Verhaegh2021,Verhaegh2023,Verhaegh2023b,Greenhouse2024}, and a schematic illustration of the (non-Bayesian version of the) analysis can be found in figure 2 of \cite{Verhaegh2021}.

It is assumed that the plasma is optically thin in all presented atomic emission analyses (see further discussion in section \ref{ch:photon_opacity}). For the MAST-U cases studied, the amount of photon opacity is expected to be negligible, as the divertor densities and power levels obtained are significantly lower than required for conditions in which photon opacity strongly impacts hydrogenic emission, according to CRETIN modelling \cite{Soukhanovskii2022}. MAST-U photon opacity investigations are planned in future work when VUV spectroscopy is available on MAST-U.

\subsection{Evolution of ion sources and sinks during detachment}
\label{ch:IonProf}

\begin{figure}
\centering
\includegraphics[width=0.8\linewidth]{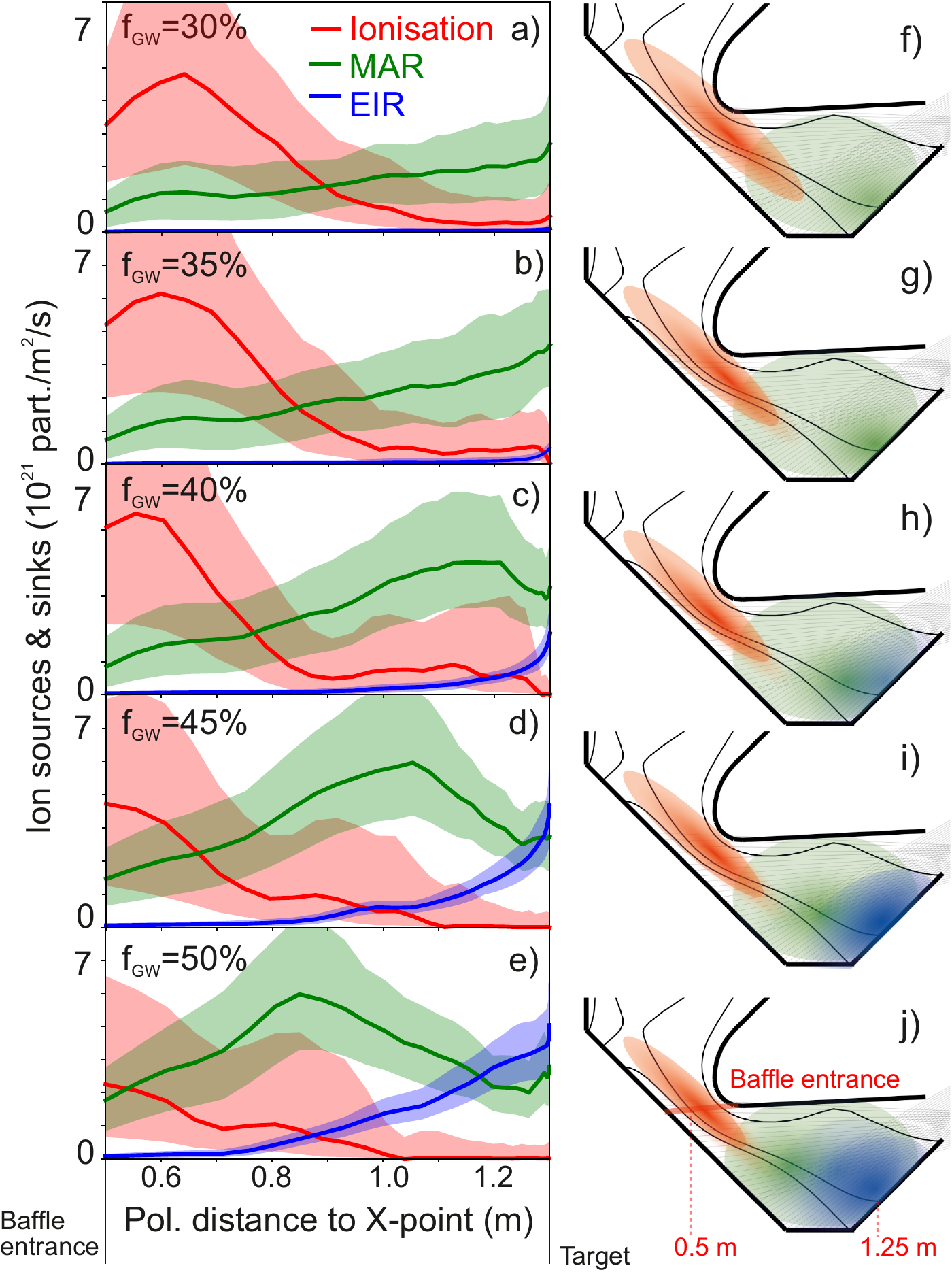}
\caption{a-e) Profiles of ion sources and sinks as function of poloidal distance to the X-point for the Super-X divertor for \# 46860 (a-e), together with a schematic illustration of the ion sources and sinks (f-i). The profiles and illustration are shown at $f_{GW} = 30 \%$ (a,f) to $f_{GW} = 50 \%$ (e,i). In a-e, 0.5 m is at the baffle entrance and 1.25 m is at the target of the Super-X divertor, as indicated in j.}
\label{fig:ProfsSXD}
\end{figure}

Ion source and sink profiles, obtained through spectroscopic analysis \cite{Verhaegh2019a,Verhaegh2023b}, are shown in figure \ref{fig:ProfsSXD} as detachment progresses during a core density scan. Comparing these L-mode beam-heated results ($P_{SOL} = 1.2$ MW) with the ionisation and recombination profiles obtained during ohmic L-mode density ramps ($P_{SOL} = 0.5$ MW) at similar core and upstream densities \cite{Verhaegh2023b,Verhaegh2023}, we observe qualitatively similar results.  

During density ramps under ohmic conditions, as detachment progressively deepens, four phases of detachment were identified in \cite{Verhaegh2023,Verhaegh2023b}, summarised in the following. 
\begin{enumerate}[label=\textnormal{(\Roman*)}]
    \item The ionisation front detaches from the target, giving rise to an increase in neutral atom and molecular density below it, giving rise to MAR and MAD. In the beam-heated L-mode case, the ionisation front is detached from the target throughout the density scan ($f_{GW} > 30 \%$, figures \ref{fig:ProfsSXD}a-e). MAR and MAD becomes stronger as the ionisation region moves further upstream.
    \item As the divertor cools further, the MAR peak detaches from the target (figure \ref{fig:ProfsSXD}c), presumably because the electrons start to have less energy to provide significant vibrational excitation ($T_e < 0.8$ eV) (see section \ref{ch:CRM_mod} for further explanation). As detachment proceeds, the MAR peak moves upstream (figures \ref{fig:ProfsSXD}d-e).
    \item As even lower electron temperatures are obtained ($T_e \ll 0.5$ eV), Electron-Ion Recombination (EIR) begins to occur (figure \ref{fig:ProfsSXD}c) near the target and the EIR region grows further upstream as the divertor detaches more deeply (figure \ref{fig:ProfsSXD}d-e).
    \item Ultimately, during the deepest detached phases under Ohmic conditions \cite{Verhaegh2023,Verhaegh2023b}, the bulk electron density moves upstream, leading to a reduction of the EIR ion sink near the target and a movement of the EIR region upstream. At this time, $T_e<0.15$ eV is observed \cite{Verhaegh2023b}. This is not observed even in the most deeply detached state for the beam-heated L-mode scenario (figure \ref{fig:ProfsSXD}e), as $T_e\ll0.2$ eV conditions are no longer achieved.
\end{enumerate}

These new NBI heated results feature double the divertor electron density (from $1-2 \times 10^{19} \textrm{m}^{-3}$ \cite{Verhaegh2023} under ohmic conditions to $2-4 \times 10^{19} \textrm{m}^{-3}$ in beam-heated conditions). This has been inferred from the Stark broadened spectra of line-of-sight spectroscopy \cite{Verhaegh2023}, divertor Thomson scattering, as well as novel coherence imaging spectroscopy to monitor the Stark broadening of the $D\gamma$ transition in 2D \cite{Lonigro2023} in the Super-X divertor. Due to the higher electron densities, EIR begins to occur at higher electron temperatures, and thus the detachment of the MAR peak from the target no longer necessarily occurs before the EIR onset (figure \ref{fig:ProfsSXD}c). 

With higher power conditions, the evolution of the ion source/sink profiles, in terms of both the magnitude and front location of the divertor ionisation source, seems to have become less sensitive to changes in the core density than in Ohmic L-mode conditions presented previously \cite{Verhaegh2023b,Verhaegh2023}. However, to solidify this observation, a more systematic study of the impact of $P_{SOL}$ on the location and magnitude of ionisation is required, using the same scenario. Additionally, such studies should account for the increase in NBI off-axis absorption at higher densities, which can increase by up to 300 kW throughout the density scan, resulting in the variation of $P_{SOL}$ shown in table \ref{tab:ShotParam}. 

\subsection{Particle balance and divertor power loss analysis}
\label{ch:PowerPart}

\begin{figure}
\centering
\includegraphics[width=0.8\linewidth]{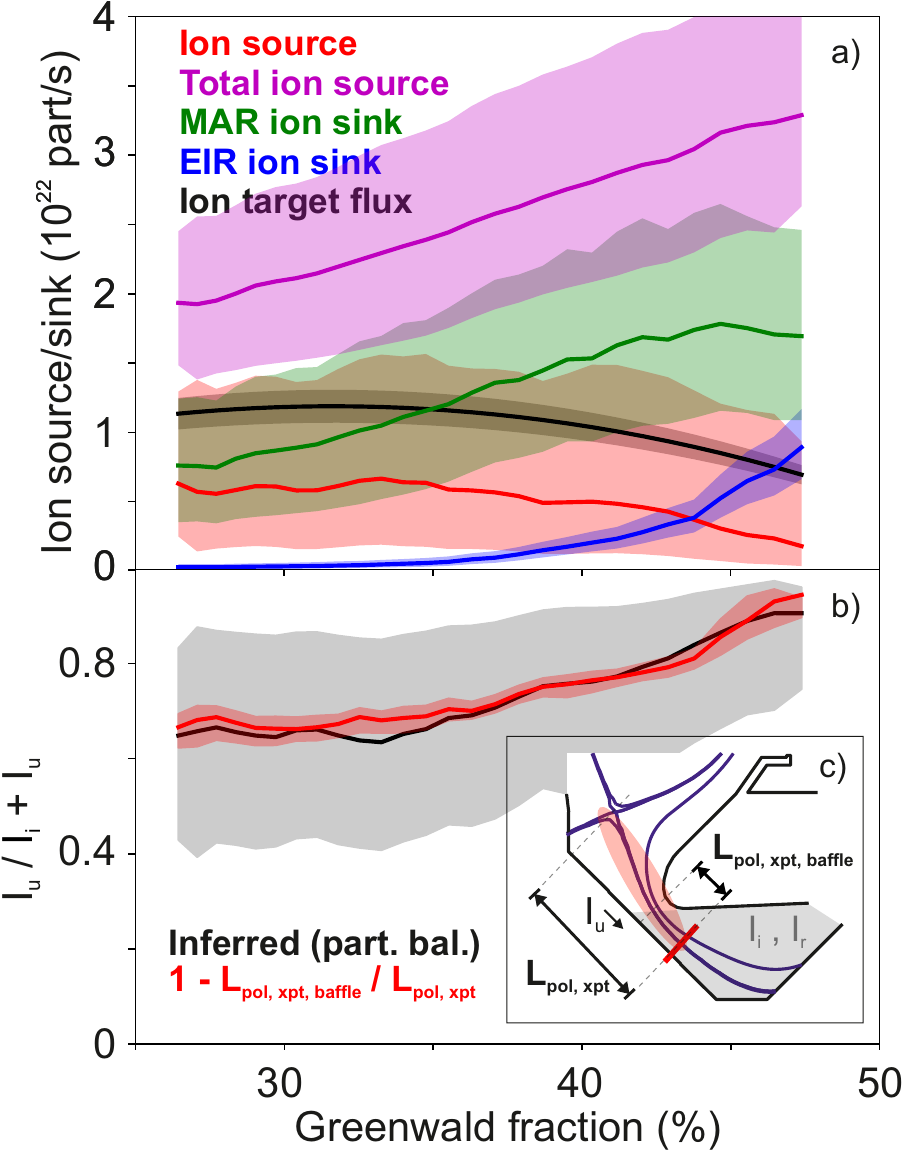}
\caption{a) Divertor particle balance for Super-X divertor as a function of the core Greenwald fraction for \# 46860, in terms of target ion flux, ion sources, MAR ion sinks, EIR ion sinks and the inferred total ion source (equation \ref{eq:PartBal}). b) Fraction of the total ion source ($I_u+I_t$) obtained from ion flows into the divertor chamber ($I_u$), inferred from particle balance (equation \ref{eq:PartBal}). This is compared against the fraction of the poloidal leg length, illustrated schematically in c, from the ionisation front to X-point ($L_{pol, xpt}$), that is above the baffle: $1 - \frac{L_{pol, xpt, baffle}}{L_{pol, xpt}}$. The grey shaded area in c) indicates the volume over which the ion source ($I_i$), MAR and EIR ion sinks ($I_r$) are obtained spectroscopically, with an indication of the inflow of $I_u$ from outside this region.}
\label{fig:PartBalSXD}
\end{figure}

Integrating the ion source and sink profiles throughout the divertor chamber shows that Molecular Activated Recombination (MAR) ion sinks substantially impact the ion target flux under NBI-heated conditions. Figure \ref{fig:PartBalSXD} shows that the MAR ion sinks within the divertor chamber (depicted in green) are of a similar order of magnitude as the sum of the total sources of atomic and molecular ionisation in the divertor chamber (illustrated in red). 

Electron-Ion Recombination can emerge as a significant ion sink mechanism, but only at the highest core electron densities. However, even during the deepest detached state obtained, the MAR ion sink downstream the baffle consistently outweighs the EIR ion sink, in agreement with the findings from previous ohmic L-mode discharges \cite{Verhaegh2023b}. The appearance of MAR (and MAD) at temperatures below 1 eV is in agreement with collisional-radiative modelling predictions using cross-sections for molecular charge exchange from Ichihara \cite{Ichihara2000} (see \cite{Verhaegh2023b,Kobussen2023} and section \ref{ch:vibr_states}, figure \ref{fig:DmD2p_MAR}). 

Although ion sinks balance ion sources from the start of discharge, the ion target flux detected by target Langmuir probes \footnote{The data shown is obtained as a polynomial fit through the Langmuir probe data from the lower and upper divertors.} is still substantial at the lowest core densities (black). Using the particle balance (equation \ref{eq:PartBal}), $I_u$ and the total ion source $I_u + I_i$ can be inferred: $I_u = I_t + I_r - I_i$ and $I_u + I_i = I_t + I_r$, respectively. The large difference between the total ion source (magenta in figure \ref{fig:PartBalSXD}a) and the divertor ion source (red) shows that $I_u$ plays a dominant role in the total ion source, particularly at the end of the discharge. Figure \ref{fig:PartBalSXD}b shows the fraction of the total ion source due to an upstream ion inflow: $\frac{I_u}{I_u + I_i} = \frac{I_t + I_r - I_i}{I_t + I_r}$, rising from 60 \% at the lowest core densities to 90 \% at the end of the discharge. 

Although the ionisation \emph{front} is relatively insensitive to changes in core density, a significant part of the ion source occurs outside the divertor chamber. However, since the ionisation front \footnote{The ionisation front is defined as the point where the line-integrated ionisation source reaches ($2 \pm 0.5 \times 10^{21} \textrm{ part/m}^2\textrm{/s}$), based on the ionisation source profiles presented in figure \ref{fig:ProfsSXD}.} is detached from the target and moves further upstream as the core density is increased, the fraction of the poloidal distance ($L_{pol,xpt}$) between the ionisation front and the X-point that is outside the divertor chamber ($1 - \frac{L_{pol, xpt, baffle}}{L_{pol,xpt}}$, figure \ref{fig:PartBalSXD}c) is large and increases as detachment progresses (figure \ref{fig:PartBalSXD}b). This fraction is in agreement with the fraction of upstream inflow of ions to the total ion source ($\frac{I_u}{I_i + I_u}$), suggesting that this magnitude of the ion inflow is expected if constant ionisation is expected between the X-point and the ionisation front without any additional ionisation upstream of the X-point (figure \ref{fig:PartBalSXD}b).

\begin{figure}
\centering
\includegraphics[width=\linewidth]{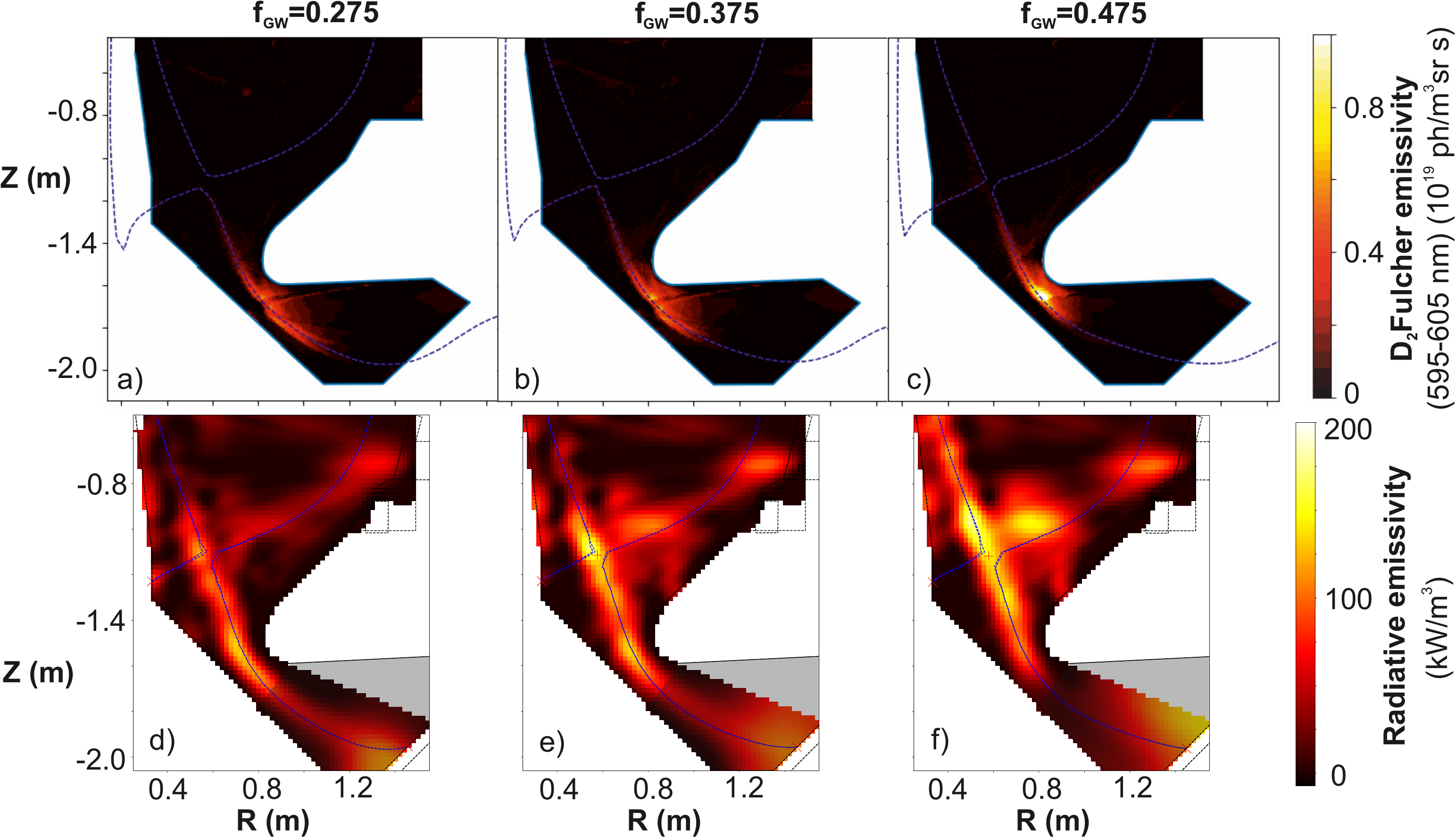}
\caption{a-c) 2D inversion of $D_2$ Fulcher emission (595-605 nm bandpass filter - \cite{Wijkamp2023}), obtained from combined inversions of the Multi-Wavelength Imaging (MWI) \cite{Wijkamp2023,Feng2021} and X-point Imaging (XPI) diagnostic at three different core densities for \# 48330, repeat discharge of \# 46860. d-f) 2D inversion of radiative emissivity, obtained from an imaging bolometer (IRVB \cite{Federici2023}) for \# 47958, repeat discharge of \# 46860. The separatrix is indicated by a blue dotted line. Due to a lack of spatial coverage only the total radiation, rather than its distribution, is reliable in the grey shaded area of figures d-f \cite{Federici2023}.}
\label{fig:D2Fulcher_XPT}
\end{figure}

Measurements from a novel X-point imaging system show intense $D_2$ Fulcher band emission between the baffle entrance and the X-point, but not upstream of the X-point, throughout the core density scan (figure \ref{fig:D2Fulcher_XPT}a-c) \footnote{An inversion artefact is present near the baffle entrance (r=0.9 m, z=-1.65 m), where there is a small gap between optimal coverage of the MWI and XPI diagnostics.}. As highlighted in previous work \cite{Verhaegh2023,Verhaegh2023b}, the presence of $D_2$ Fulcher emission indicates the presence of electron impact dissociation, which can be used as a proxy for the presence of an ionising plasma. Therefore, these measurements support the view that ionisation between the baffle entrance and the X point can lead to a significant $I_u$ \footnote{It should be noted, however, that low atomic hydrogen density at the separatrix can still result in a significant ionisation source, due to the large volume integral. For these MAST-U conditions, the fuelling is directed at the low-field side separatrix, and the fuelling flux (table \ref{tab:ShotParam}) is non-negligible compared to the total estimated ion source. Information on this can be obtained by comparing the total ionisation along the entire divertor leg (obtained from the quantitative analysis of the X-point imaging data) to the inferred total ion source, which is future work.}. This information is qualitatively consistent with measurements from an imaging bolometer \cite{Federici2023} (figure \ref{fig:D2Fulcher_XPT}d-f), which indicates significant radiative losses: 1) in between the X-point and the baffle entrance; 2) at the region where $D_2$ Fulcher emission occurs. 

\begin{figure}[htbp]
\centering
\includegraphics[width=0.6\linewidth]{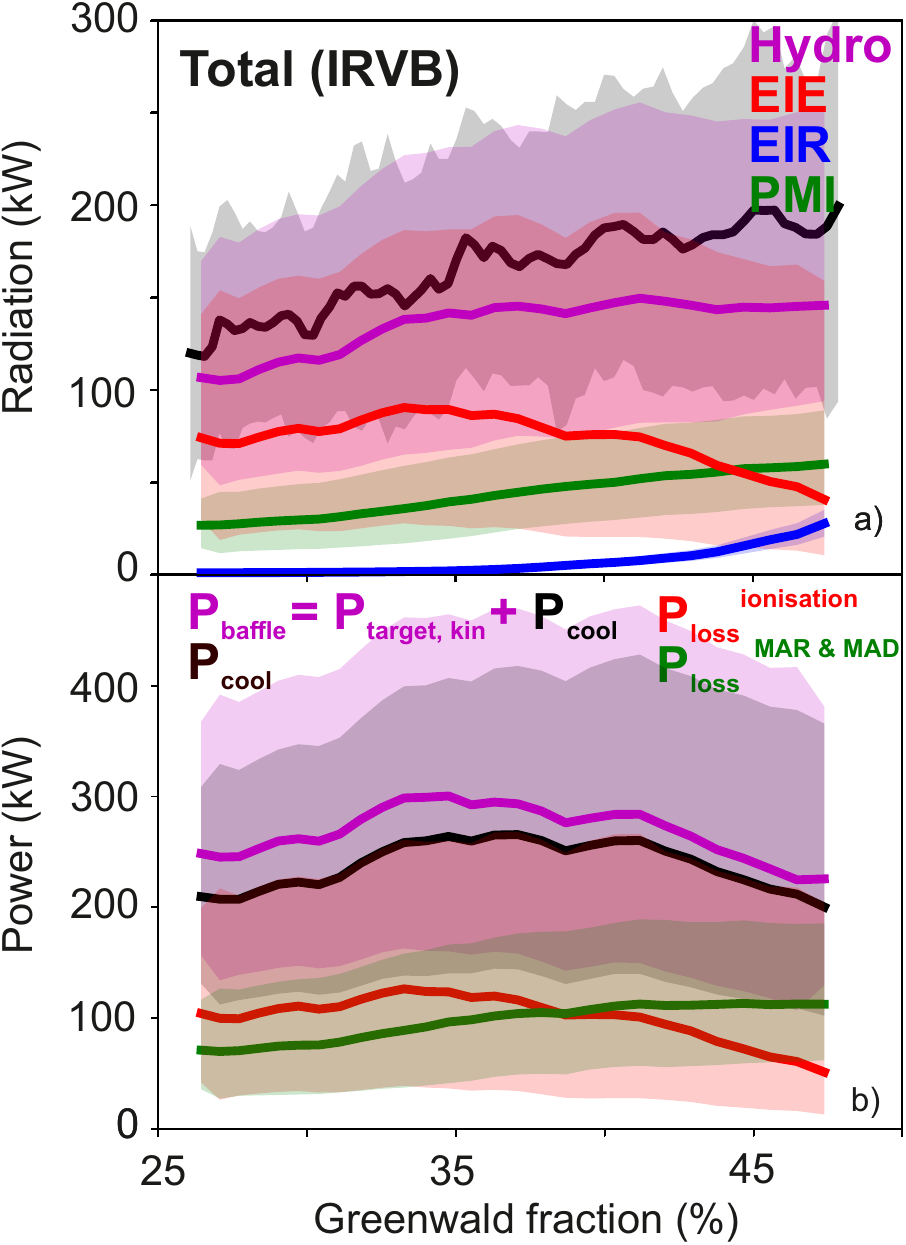}
\caption{a) Radiative power losses integrated downstream the baffle entrance for \# 46860. Black - total radiation (imaging bolometer - IRVB \cite{Federici2023} within the shaded region of figure \ref{fig:PartBalSXD}c), from repeat discharge \# 47958. Total atomic hydrogenic radiation ('Hydro' - magenta) consisting of radiation from electron impact excitation ('EIE' - red), electron-ion recombination ('EIR' - blue) and plasma-molecular interactions ('PMI' - green), inferred from spectroscopic analysis. b) Estimated power input ($P_{baffle}$ magenta) at the baffle entrance, consisting of total hydrogenic electron cooling from the divertor ($P_{cool}$ - black) plus the target power load ($P_{target, kin}$). $P_{cool}$ mainly consists out of power losses from ionisation ($P_{cool}^{ionisation}$ - red) and power losses from MAD and MAR ($P_{loss}^{MAR \& MAD}$ - green).}
\label{fig:PowerLossSXD}
\end{figure}

The various hydrogenic reactions in the divertor have an impact on the divertor particle balance and lead to divertor power losses. The total inferred hydrogenic divertor radiative power losses, integrated over the divertor chamber\footnote{These have been monitored in the lower divertor and have been multiplied by two under the assumption that the upper/lower divertors are symmetric, in agreement with the measured upper/lower divertor ion target fluxes.}, are similar to the total radiation monitored by an imaging bolometer (IRVB) \cite{Federici2023} in that region (figure \ref{fig:PowerLossSXD}). These higher power findings ($P_{SOL} \sim 1.2$ MW) are consistent with previous ohmic findings ($P_{SOL} \sim 0.5-0.6$ MW) \cite{Verhaegh2023b}. However, in contrast to previous ohmic findings, the majority of this hydrogenic power loss is attributed to electron-impact excitation, which is sustained throughout the discharge as the ionisation source, downstream the baffle, remains significant throughout the density scan. 

\setcounter{footnote}{0}

However, the total net electron cooling is attributed to more processes than just volumetric radiation. To dissociate a molecule and ionise the neutral atoms created, energy \footnote{However, the electron cooling associated with these energies can be (partially) delivered to the target upon surface recombination (and re-association), therefore the net electron cooling is not necessarily a net power loss.} is required. Taking this into account, the total net electron cooling \footnote{This assumes that energy gains from MAR and EIR are ultimately re-delivered to the electrons through equilibration.} from hydrogenic processes, downstream of the baffle, is shown in figure \ref{fig:PowerLossSXD}b. This shows that the dominant contributors to electron cooling are ionisation and MAD, which can increase to 20 \% of $P_{SOL}$ (1.0-1.3 MW). Combining the hydrogenic radiative power loss from the divertor chamber with estimated target power loads \footnote{The target power load is inferred by combining the measured ion target flux (figure \ref{fig:PowerLossSXD}a) with inferred target temperatures from spectroscopic analysis: $P_{target, kin} = \gamma I_t T_t$, assuming $\gamma=7$ and neglecting surface recombination to avoid double counting.} enables inferring the total power input at the baffle entrance: $P_{baffle} = 300 \pm 150$ kW (figure \ref{fig:PowerLossSXD}). This power input is small compared to the estimated $P_{SOL}$. However, the particle balance analysis (figure \ref{fig:PartBalSXD}), shows the ion source upstream of the baffle entrance ($I_u$) is significant. Electron cooling associated with $I_u$ can be up to 385 kW (assuming a net ionisation energy loss of up to 30 eV) and could partially explain the relatively minor power inputs into the divertor chamber, which are still significantly larger than those obtained under ohmic conditions \cite{Verhaegh2023,Moulton2023}.

Significant power losses upstream of the baffle entrance, up to the X-point are observed with imaging bolometry (figure \ref{fig:D2Fulcher_XPT}d-f). Some of these radiative losses occur upstream of the $D_2$ Fulcher emission (and thus ionisation) region (figure \ref{fig:D2Fulcher_XPT}a-c), suggesting that impurity radiation can play a role upstream of the baffle, consistent with core VUV spectroscopy measurements \cite{Verhaegh2023b}.

In summary, our investigation into the evolution of ion sources and sinks during detachment in the MAST-U Super-X divertor under beam-heated conditions has revealed a behaviour qualitatively similar to previous ohmic results. However, the ionisation, downstream the baffle entrance, remains sustained throughout the density ramp, leading to sustained power losses. Despite these higher power conditions, MAR and MAD continue to have, respectively, dominant and significant impacts on divertor particle and power balance; suggesting MAR and MAD can be significant in higher power conditions. 

\subsection{Impact of additional heating on the detachment state}
\label{ch:PNBI}

\begin{figure}[htbp]
\centering
\includegraphics[width=\linewidth]{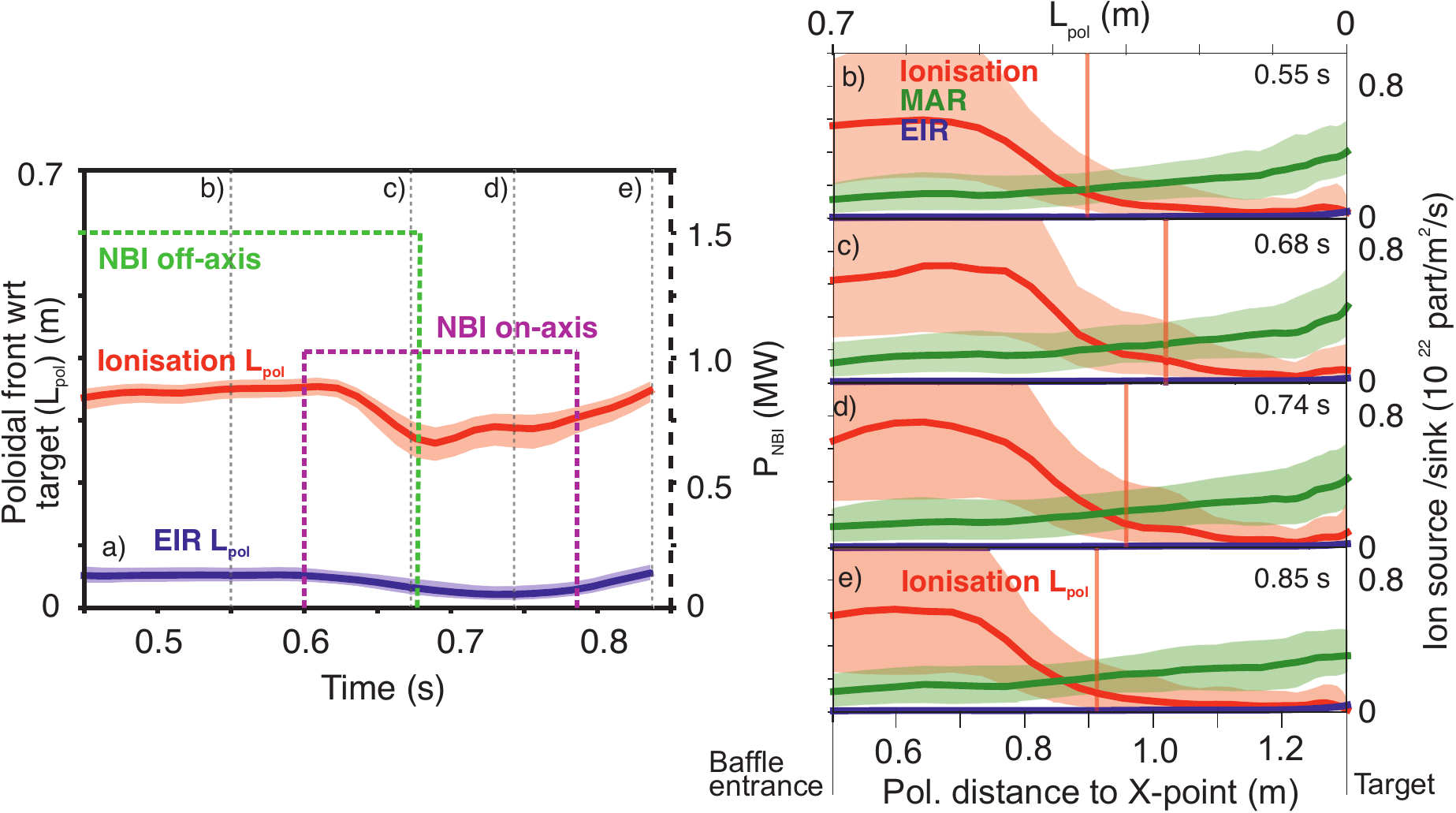}
\caption{a) Evolution of the ionisation and electron-ion recombination front poloidal position \emph{with respect to the target as opposed to the X-point} - $L_{pol}$ as function of time, for discharge \# 46705. The dotted trends indicate the levels of NBI power used. The vertical dotted lines indicate the time points at which the ion source/sink profiles are shown in figures b-e). b-e) Ion source and sink profiles of the total ion source (red), MAR ion sink (green), and EIR ion sink (blue) as function of poloidal distance to the X-point and target ($L_{pol}$). The red vertical lines indicate the ionisation front position shown in a.}
\label{fig:PowerTransientSXD}
\end{figure}

Our results show that the NBI heated results (1.5 MW off-axis) of the Super-X divertor remain qualitatively consistent with previous ohmic findings. Likewise, the behaviour of the divertor remains qualitatively similar in plasmas with 1 MW additional on-axis NBI heating, doubling the absorbed NBI power (table \ref{tab:ShotParam}), whilst the core density is kept constant at $f_{GW} = 30 \%$ (figure \ref{fig:PowerTransientSXD}) \footnote{The total estimated NBI beam-absorption is $\sim 0.8$ MW during the off-axis NBI heated phase, increasing to $\sim 1.7$ MW when both beams are introduced. Although the core radiation changes negligibly during the dual beam-heated phase and remains below 250 kW, the change in $P_{SOL}$ cannot be accurately estimated as the plasma stored energy changes when on-axis NBI is added.}. The doubling of absorbed beam power only results in a small movement of the ionisation and EIR fronts \footnote{The reference values for the ionisation / EIR front are defined as $(1.5 \pm 0.25) \times 10^{21} \textrm{ part./m}^2\textrm{/s}$ and $(1.75 \pm 0.25 \times 10^{20}) \textrm{ part./m}^2\textrm{/s}$ respectively.} by $\sim 10$ cm and $\sim 3$ cm respectively (compared to a 1.25 m divertor poloidal leg length, 0.7 m of which is in the divertor chamber) (figure \ref{fig:PowerTransientSXD}a). The ion source and sink profiles (figure \ref{fig:PowerTransientSXD}b-e) only change slightly during the power scan: leading to a total increase of the divertor ion source by $\sim 25 \%$, while the total ion sink increases by $\sim 15 \%$ (figure \ref{fig:PowerTransientSXD}b-e). 

This suggests that the MAST-U Super-X results extend to even higher power conditions. The observation that additional heating results in only a minor impact on the detached state and that it results in additional ion sinks showcases the Super-X's capability for enhanced particle exhaust. Additionally, the insensitivity of the ionisation region to changes in heating is consistent with the observed insensitivity of the ionisation region to changes in density (figure \ref{fig:ProfsSXD}: $\sim 10$ cm movement of the ionisation front as the core density increases from 30 to 50 \% of the Greenwald fraction). 

\section{Discussion}

Our results highlight that the promising initial results of the MAST-U Super-X divertor \cite{Verhaegh2023b}, extend to higher power conditions. In particular, plasma-molecular interactions leading to MAR and MAD remain important at higher power in MAST-U. 

$D^-$ is generated by dissociative attachment ($e^- + D_2 \rightarrow D_2^- \rightarrow D^- + D$) and results in MAR ($D^+ + D^- \rightarrow D + D$), MAD ($D^- + D^+ \rightarrow e^- + D^+ + D$) and atomic line emission from the excited atoms generated in these processes. Analysing the emission from these excited atoms shows that a potential occurrence of such interactions is consistent with our measurements and with new ab initio electron attachment cross-section calculations for deuterium \cite{Laporta2021}. However, $D_2^+$ remains the dominant MAR (and MAD) process. 

One question that remains is whether MAR and MAD can be relevant in reactors. By post-processing reactor relevant simulations that feature tightly baffled long-legged divertors, we show that such interactions can have dominant impacts. We discuss the implications of these findings for diagnostic development and detachment control, as well as for the use of exhaust simulations for reactor design. 

\subsection{The presence of $D^-$ and its impact on particle balance}
\label{ch:MAR_MAD_Dm}

\begin{figure}[htbp]
\centering
\includegraphics[width=0.5\linewidth]{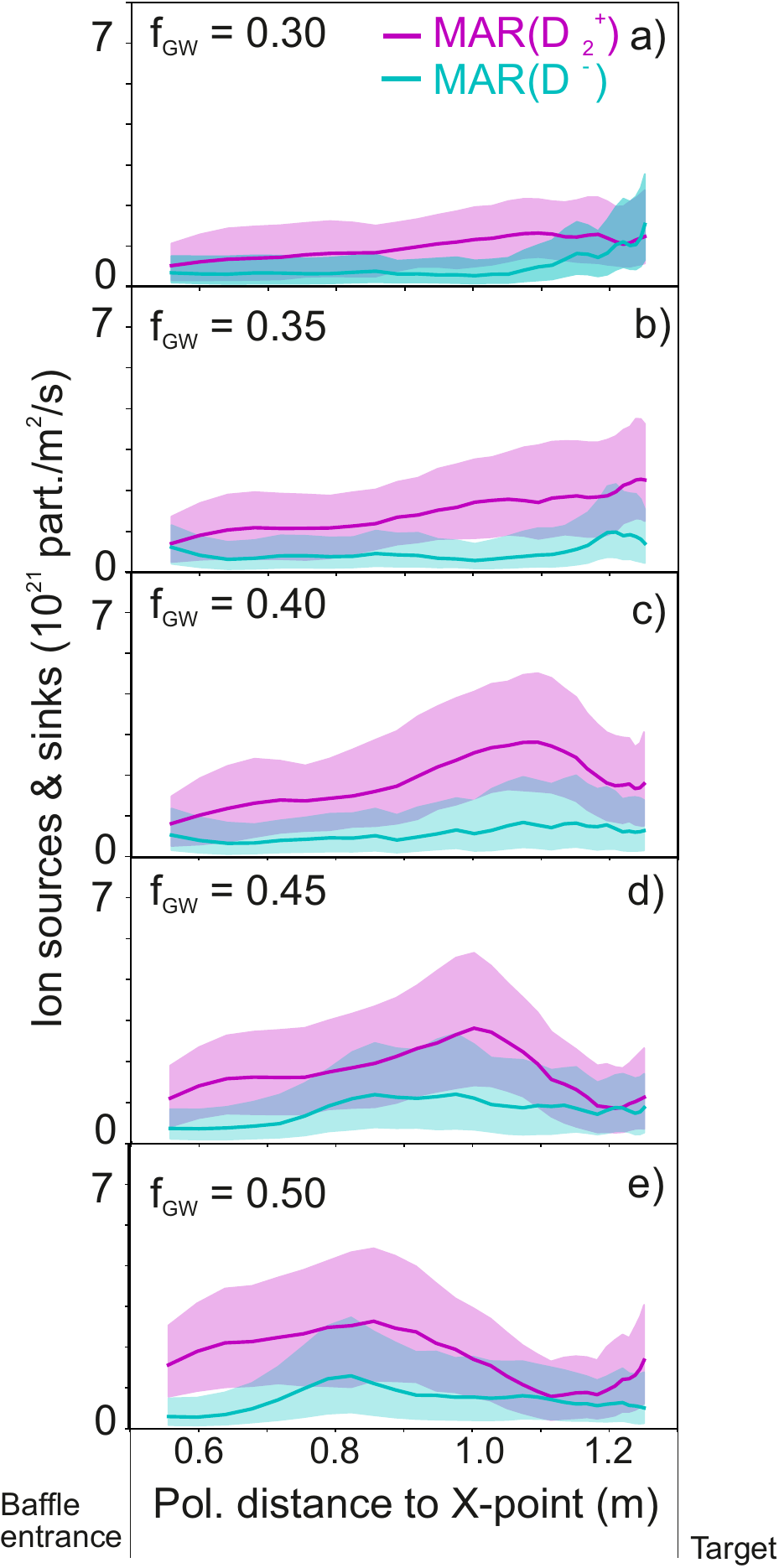}
\caption{a-e) Profiles of inferred MAR ion sinks associated with $D_2^+$ and $D^-$ as function of poloidal distance to the X-point for the Super-X divertor at different core densities for \# 46860.}
\label{fig:MAR_Dm}
\end{figure}

The existence of significant $D^-$ in the divertor has been extensively debated in the literature \cite{Kukushkin2017,Reiter2018,Verhaegh2021}, as measurements (in the vibrational ground state) indicate that the electron attachment cross-section is much smaller for deuterium than for hydrogen \cite{Krishnakumar2011}. However, the isotope dependency is strongly reduced at higher vibrational levels, where the cross-section for electron attachment is also enhanced, causing uncertainties over whether $D^-$ could or could not be relevant. Balmer line emission analysis, on the TCV divertor using $D\alpha, D\beta$ and two other medium-n Balmer line measurements, suggested that $D^-$ may indeed contribute significantly to MAR and MAD \cite{Verhaegh2021b}. Plasma spectroscopy in linear devices, operating with hydrogen plasmas that have divertor-relevant plasma parameters, indicate $H^-$ can have a strong impact on MAR and MAD \cite{Federici2024}, 

Using new data from MAST-U, where diagnostic repeat discharges were performed to obtain $D\beta$ emission measurements, the potential presence and impact of $D^-$ are inferred for the cases presented in section \ref{ch:PartPowerBal} using a Bayesian version of the analysis developed in \cite{Verhaegh2021}. This uses the same procedure described previously (section \ref{ch:PartPowerBal}), but adds an additional free parameter that separates the $D\alpha$ emission arising from plasma interactions with $D_2^+$ and $D^-$. Since this analysis only depends on the hydrogen Balmer line emission (arising from interactions between $D^-$ and $D^+$), no explicit assumptions on dissociative attachment rates, nor the rovibronic distribution of $D_2$, are assumed \cite{Verhaegh2021} \footnote{Population coefficients for $H_2^+$ and $H^-$ from Yacora (on the Web) are used for this inference \cite{Wunderlich2020,Wuenderlich2016}, under the assumptions that they are the same for hydrogen and deuterium. To convert the emission of $D\alpha$ associated with $D^-$ into estimates of MAR and MAD, the reaction rates for $H^- + H^+ \rightarrow H + H$ and $H^- + H^+ \rightarrow e^- + H^+ + H$ have been adopted from Eirene 'AMJUEL' \cite{Reiter2005}, which was derived using the collisional-radiative model of \cite{Sawada1995}. For further information, see \cite{Verhaegh2021}.}. 

The results in figure \ref{fig:MAR_Dm} shows $D^-$ could indeed be present and contribute to hydrogen emission, MAR ion sinks, and MAD, consistent with previous TCV findings \cite{Verhaegh2021b}. However, the impact of $D^-$ is expected to be smaller than that of $D_2^+$ and its uncertainty is substantial. From our statistical analysis we find that there is a 68 \% confidence that the contribution of $D^-$ to the total MAR is at least 20 \%. Although $D\beta$ measurements provide additional information to constrain the $D^-$ content, there are sufficient other degrees of freedom in our emission model to make $D^-$ imperfectly constrained without additional measurements (emission pathlengths, electron temperatures, electron densities, neutral atom densities, see \cite{Verhaegh2021}).

\subsection{MAR and MAD rates of $D^-$ and $D_2^+$}
\label{ch:CRM_mod}

State-of-the-art implementation of exhaust modelling codes, such as SOLPS-ITER, EDGE2D-EIRENE and EMC3-EIRENE, generally omit the electronic, vibrational, and rotational splitting of the molecules and only evaluate transport of ground-state molecules. Effective rates are used that act on those ground-state molecules, computed by collisional-radiative models which assume that the vibrational distribution of the molecules reaches a quasi-steady state based on local plasma parameters. 

Dissociative attachment is generally ignored in SOLPS-ITER and previous work has highlighted errors in the molecular charge exchange rates employed by Eirene \cite{Verhaegh2021b,Verhaegh2023a,Verhaegh2023b}. To further investigate this, we calculate new effective molecular charge exchange rates in this section, including their impact on MAR and MAD, using vibrationally resolved rates from Ichihara \cite{Ichihara2000}, for molecular charge exchange, and Laporta \cite{Laporta2021}, for dissociative attachment (assuming rotational and electronic ground state). Both the Ichihara/Laporta rates/cross-sections are ab initio generated for every vibrational level and thus do not rely on rescalings applied to the vibrational ground state, as is the case with the cross-sections used by Eirene \cite{Greenland2001}. 

Remaining as close as possible to the Eirene implementation, these rates are calculated analogously using a simplified collisional-radiative model (CRM) \cite{Kobussen2023,Verhaegh2023b,Holm2022} to compute the vibrational distribution ($f_\nu (T_e=T_i=T)$), which is applied as a weighting factor on vibrationally resolved rates to compute effective rates. In addition to updated dissociative attachment and molecular charge exchange data \footnote{It should be noted that whether electron-attachment is included or not has a negligible impact on the effective molecular charge exchange rate.}, this uses the same approach and rates as used in the calculation of effective rates in Eirene. It assumes electronic and rotational ground state and ignores any coupling between rotational, vibrational, and electronic levels. This approach accounts for any depletion of vibrationally excited molecules due to the increased dissociative attachment/molecular charge exchange rates self-consistently. More information can be found in \ref{ch:colrad}.

\begin{figure}[H]
\centering
\includegraphics[width=\linewidth]{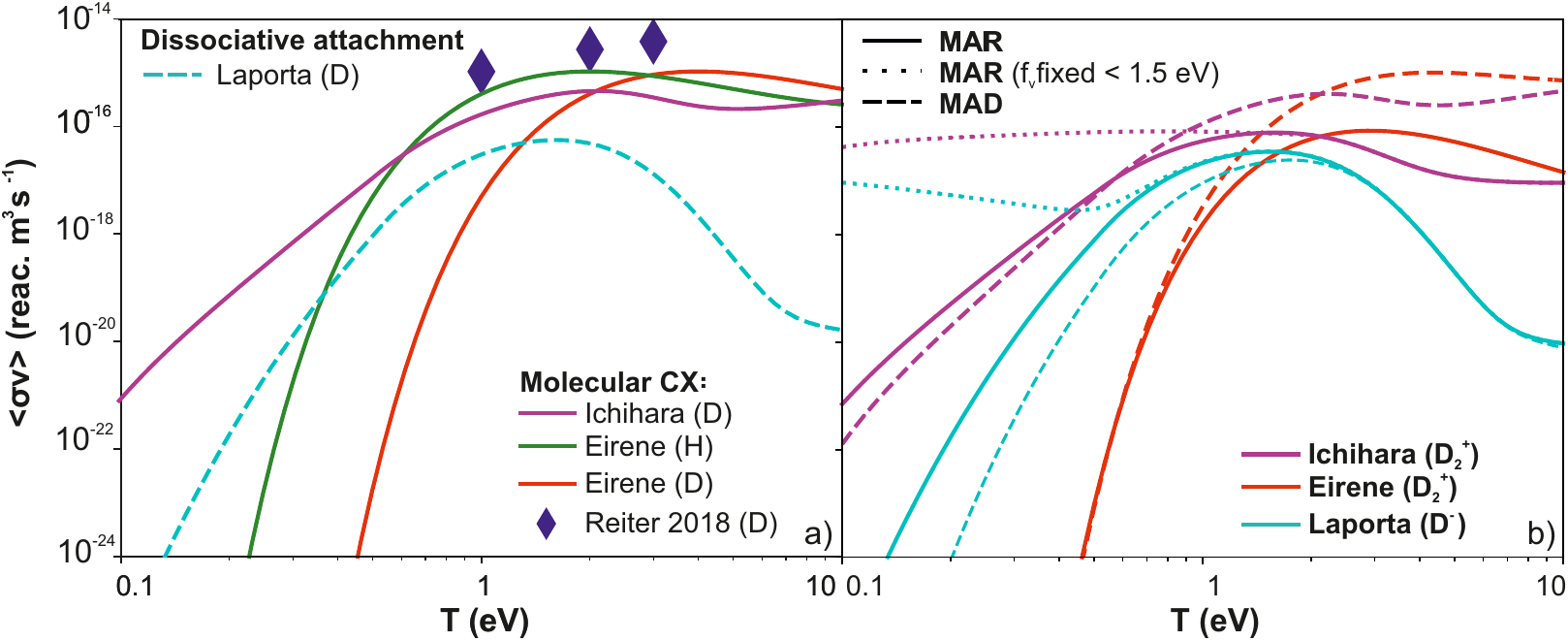}
\caption{a) Effective molecular charge exchange (Ichihara - magenta - solid) and dissociative attachment rates (Laporta - cyan - dashed) as function of temperature ($T=T_e=T_i$) for deuterium, assuming $n_e = 5 \cdot 10^{19} m^{-3}$ (b). For comparison, the effective molecular charge exchange rates assumed by Eirene (for hydrogen (green - solid) and deuterium (red - solid)) are also shown \cite{Reiter2005}, as well as the effective molecular charge exchange rate calculated in \cite{Reiter2018} for $T=1,2,3$ eV (blue symbols) 'Reiter 2018 (D)'. b) Effective MAR (solid lines) and MAD (dashed lines) rates for $D_2^+$ and $D^-$, using the molecular charge exchange and dissociative attachment rates from a). To highlight the sensitivity of the effective rates to $f_\nu$, recalculated MAR rates (dotted lines) are shown assuming the vibrational distribution remains unaltered below 1.5 eV ($f_\nu (T<1.5 eV) = f_\nu (T=1.5 eV)$).}
\label{fig:DmD2p_MAR}
\end{figure}

The effective dissociative attachment rate (D) and the molecular charge exchange rate (D) \cite{Verhaegh2023b} are shown in figure \ref{fig:DmD2p_MAR}a, together with the Eirene molecular charge exchange rate (H and D). These new rates, in contrast to those employed in Eirene, predict that MAR and MAD are sustained at low temperatures $T_e<1$ eV, consistent with experimental results \cite{Verhaegh2023,Verhaegh2023b}. MAR arising from $D^-$ could indeed be significant near 1 eV (figure \ref{fig:DmD2p_MAR}), however MAR from $D^-$ is expected to be less likely at low temperatures, despite the electron-attachment cross-sections being increased at such low temperatures, because the electrons have less energy to cause vibrational excitation to high levels  \footnote{It should be noted, however, that this calculation assumes that $D^-$ only reacts with $D^+$. Other interactions that can deplete $D^-$ (such as neutral collisions) have not been accounted for and require further study \cite{Bergmayr2023}.}. This makes MAR (and MAD) arising from $D^-$ even more sensitive to the vibrational distribution than $D_2^+$. 

\subsubsection{Caveats, uncertainties and discussion on the calculated MAR and MAD rates}
\label{ch:vibr_states}

The mechanisms leading to molecular charge exchange and dissociative attachment and their impact on the divertor state are complex, leading to various unknowns and uncertainties that require further study. 

Both dissociative attachment and molecular charge exchange are strongly dependent on the vibrational distribution, which is uncertain \cite{Verhaegh2023b}. However, the magnitude of the newly calculated effective rates is more likely to be underestimated than overestimated. To illustrate this, figure \ref{fig:DmD2p_MAR}a compares the effective molecular charge exchange ('Ichihara (D)') against predictions from \cite{Reiter2018} (for deuterium) at $T=1,2,3$ eV ('Reiter 2018'). This uses the same Ichihara molecular charge exchange cross-sections \cite{Ichihara2000}, but assumes 'a certain (fixed) vibrational level (Boltzmann) distribution' \cite{Reiter2018}  \footnote{The precise $f_\nu$ used has not been specified in \cite{Reiter2018}, but has been estimated to correspond to a vibrational temperature between 5000 to 10000 K in order to match the molecular charge exchange rate magnitudes provided in \cite{Reiter2018}.}, resulting in a factor 5-10 increase in the effective molecular charge exchange rate compared to the 'Ichihara (D)' calculation. Additionally, the effective rate calculation assumes that the molecules are static, which may underestimate the molecular charge exchange rate at low temperatures \cite{Ichihara2000}, given that rotational temperatures of 0.4 - 0.7 eV are detected in the MAST-U Super-X divertor \cite{Osborne2023}.

The method used here and that used by Eirene to compute the effective rates is simplified. Various reaction mechanisms are not included: neutral collisions \cite{Bergmayr2023}, ion collisions (apart from molecular charge exchange), and vibrational-vibrational exchange \cite{Krasheninnikov1997}, which require further study. Currently, the CRM results in Eirene are only valid for hydrogen, and these rates are assumed to apply to heavier isotopes: extensions are required for different isotopologues \cite{Scarlett2021}. By ignoring rovibronic coupling, the increase of vibrational excitation through electronic excitation and subsequent decay is not accounted for \cite{Chandra2023}. More complex CRMs \cite{Sawada2016,Yoneda2023} suggest that a coupling between rotational and vibrational distributions is required. Other research shows that the rotational distribution can also impact plasma-molecular interaction rates, particularly for dissociative attachment \cite{Fabrikant2002}. However, a full rovibronic resolved CRM is far more advanced than the rate calculations employed in the current state-of-the-art exhaust simulations used to model fusion plasmas, such as Eirene \cite{Reiter2005}. 

Some assumptions are inherent to the effective rate approach. Effective rate calculations assume that all vibrational levels reach quasi-steady state, ignoring the impact of plasma-surface interactions and molecular transport on the vibrational distribution. This assumption can be invalid below 1-2 eV as the time it takes for the vibrational distribution to equilibrate (when the CRM is solved time-dependently) is longer than the transport time of the molecules \cite{Kobussen2023}. This may imply that the excited levels of vibrationally resolved molecules must be tracked in exhaust simulations \cite{Wischmeier2005}. This would greatly increase the complexity of exhaust simulations, which is likely unfeasible for reactor-sized devices. Further studies on such setups are ongoing at MAST-U.

To visualise the impact of the vibrational distribution and the reduction of vibrational excitation at $T<1.5$ eV on MAR and MAD, an alternative calculation of MAR and MAD (dotted curves) is shown where the vibrational distribution remains static at temperatures below 1.5 eV (e.g. $f_\nu (T<1.5 eV) = f_\nu (f=1.5 eV)$) (see figure \ref{fig:fD2v}). This results in an increase in the effective MAR rates of $D^-$ and $D_2^+$ by $>10^{7}$ and $<10^{4}$ at 0.1 eV, respectively. This highlights that transport of vibrationally excited molecules from a relatively hot plasma ($T>1.5$ eV) into a colder plasma ($T<1.5$ eV) could potentially increase the MAR/MAD rate significantly, resulting in significant uncertainties.

\subsection{Investigating impact of molecular charge exchange in reactor-relevant conditions}
\label{ch:postproc}

\setcounter{footnote}{0}

Although there are large uncertainties associated with the plasma chemistry interactions in detached divertors, the molecular charge exchange rate presented in section \ref{ch:CRM_mod} presents a significant improvement, with provenance, from the currently used rates in Eirene that effectively disable the impact of molecular charge exchange on the divertor for deuterium plasmas \cite{Verhaegh2021b,Verhaegh2023b}, in contrast to a growing set of experimental data on multiple tokamak divertors \cite{Wijkamp2023,Verhaegh2023,Verhaegh2023a,Verhaegh2023b,Verhaegh2021,Karhunen2023,Perek2022,Hollmann2006}.

To illustrate the impact of possible underestimations of the molecular charge exchange rate, existing predictive MAST-U \cite{Myatra2023,Myatra2023a} and interpretive TCV \cite{Fil2017} SOLPS-ITER simulations were post-processed with an increased molecular charge exchange rate at $T<1.5$ eV \cite{Verhaegh2021,Verhaegh2021b} \footnote{In the absence of the CRM calculations in section \ref{ch:CRM_mod}, this previous work artificially raised the molecular charge exchange rate by turning off ion isotope mass rescaling, using the 'Eirene (H)' rate rather than the 'Eirene (D)' rate from figure \ref{fig:DmD2p_MAR}a.}. This resulted in an increase in MAR, MAD, and $D\alpha$ emission, qualitatively consistent with the experiment.

One important question that remains is whether such plasma-molecular chemistry processes can play an important role in the divertors of reactor class devices. Post-processing converged simulations of reactor-class devices with improved effective rates can be a practical important tool in answering this question. 

A schematic illustration of this post-processing approach, adopted from \cite{Verhaegh2021b,Verhaegh2021}, is highlighted in figure \ref{fig:STEP_MAR_MAD_postproc}d, which is applied after a converged simulation result is obtained (black block in figure \ref{fig:STEP_MAR_MAD_postproc}d). Neglecting the transport of $D_2^+$ and $D^-$ (which is the default of Eirene), the ratios $D_2^+ / D_2$ and $D^- / D_2$ can be approximated as the ratio between the sum of the rates that generate $D_2^+$ and $D^-$, divided by the sum of the rates that destroy them \cite{Verhaegh2021b} (green blocks figure \ref{fig:STEP_MAR_MAD_postproc}d). By multiplying these ratios (dependent on $n_e$, $T_e$) with the molecular density for each SOLPS-ITER grid cell, the densities of $D_2^+$ and $D^-$ can be inferred (figure \ref{fig:STEP_MAR_MAD_postproc}d). Using rates for $e^- + D_2^+ \rightarrow ...$ and $D^+ + D^- \rightarrow ...$, the MAR/MAD rates, emission and radiative power losses from excited atoms after plasma-molecular interactions, as well as power losses from dissociation through MAD, can be calculated from these post-processed $D_2^+$ and $D^-$ densities. By changing the molecular charge exchange rate (figure \ref{fig:DmD2p_MAR}a) in this calculation from the default Eirene rate ('Eirene (D)') to the 'Ichihara (D)' rate, and by adding dissociative attachment ('Laporta (D)'), changes in MAR, MAD, etc. can be post-processed. The extent of the increase in $D_2^+$ due to post-processing can be seen by comparing the 'Ichihara (D)' and 'Eirene (D)' rates in figure \ref{fig:DmD2p_MAR}a. For further details on the implementation, see \cite{Verhaegh2021b,Verhaegh2021}.

The advantage of post-processing is that it can be readily employed to a converged exhaust simulation without requiring new simulations. However, its disadvantage is that it is not self-consistent (as it uses rates that are different than those used in the simulations) and can thus only be used to illustrate whether MAR and MAD could potentially affect the divertor solution. New (self-consistent) simulations with modified rates are required to gauge their impacts in more detail, if MAR and MAD is significant. Following the TCV findings that the $D_2^+$ content is underestimated, such new SOLPS-ITER simulations (with the 'Eirene (H)' rate) were performed, leading to an improved agreement between experiments and simulations and retrieving the ion target flux roll-over during detachment \cite{Verhaegh2023a,Williams2022}, which was not obtained with the default rates.


Plasma-molecular chemistry would generally be more dominant in tightly baffled reactor designs. Therefore, to illustrate whether MAR and MAD associated with $D_2^+$ could play a role in reactor-class devices, SOLPS-ITER simulations of STEP \cite{Hudoba2023,Osawa2023} were used as an illustration for a post-processed reactor-class simulation \footnote{However, such post-processing can be readily employed to any SOLPS-ITER simulation.}. STEP is a reactor class design based on the spherical tokamak, featuring a double null divertor with tightly baffled inner and outer divertor legs \cite{Wilson2020,Hudoba2023,Osawa2023}. The STEP simulation used features:

\begin{itemize}
    \item High power input ($P_{SOL} \approx 80$ MW).
    \item High divertor electron densities ($\sim 10^{21} \textrm{m}^{-3}$).
    \item Impurity (argon) seeding and deuterium fuelling.
    \item Tightly baffled double null divertor geometry with an extended outer target divertor with large total flux expansion ($B_{tot, xpt}/B_{tot, t} \approx 1.8$) and an X-divertor at the inner target to improve power exhaust \cite{Hudoba2023}.
    \item The post-processed case is near the onset of detachment (ionisation front near the target).
\end{itemize}

However, a variety of other STEP simulations \cite{Hudoba2023,Osawa2023} were post-processed (not shown), leading to qualitatively similar conclusions; although the magnitude of the impact of MAR and MAD, after post-processing, varies between different cases. These simulations ranged from near detachment onset (ionisation front near target) to more deeply detached simulations, as well as simulations that included a vertical inner target divertor \cite{Osawa2023}. 

\subsubsection{The impact of molecular charge exchange on the divertor solution}

\begin{figure}
\centering
\includegraphics[width=\linewidth]{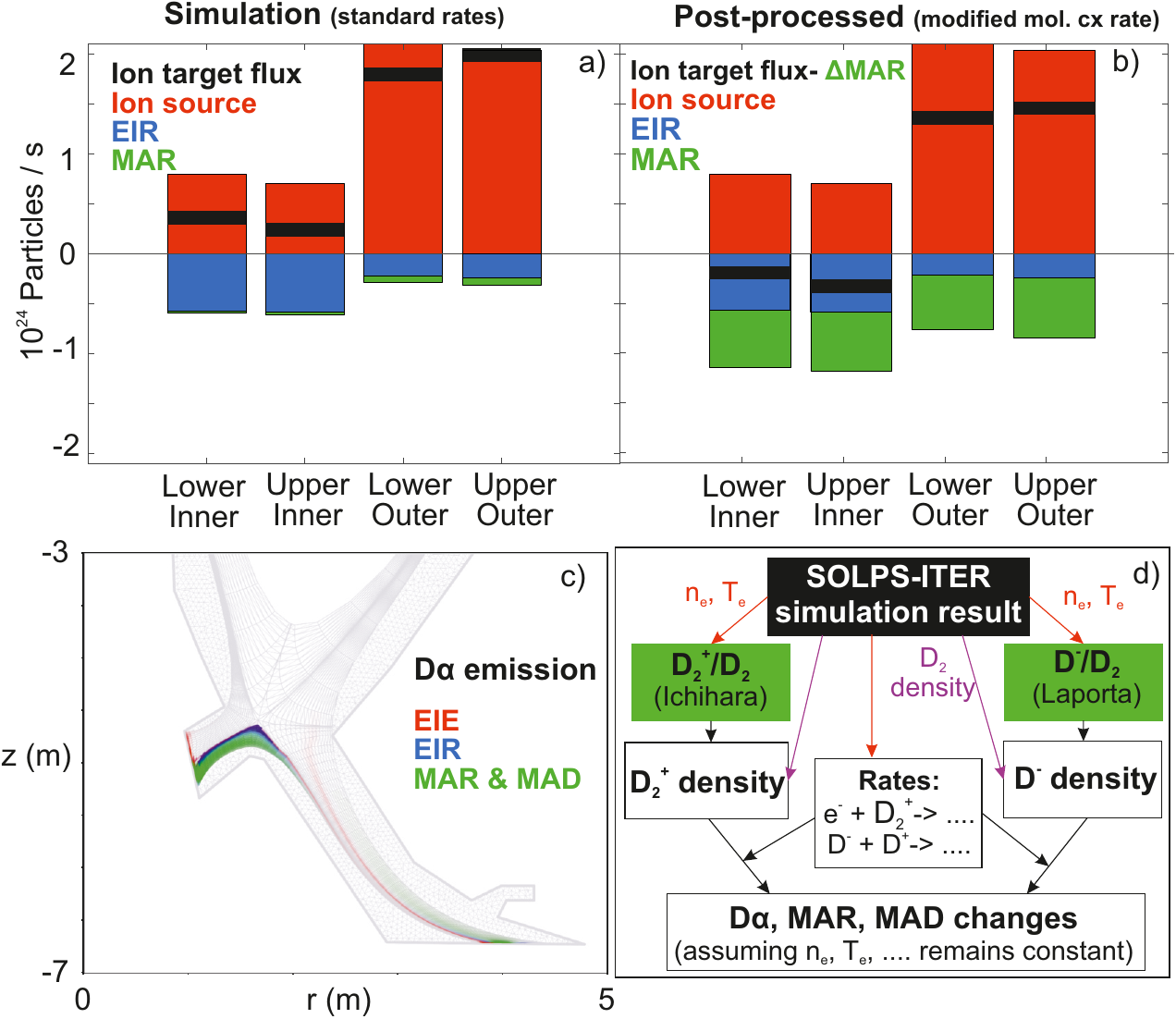}
\caption{Impact of post-processing of STEP simulations on particle (e.g., ion) balance (a,b) and $D\alpha$ emission (c). The bar-graphs illustrate the total ion sources ($>0$) and ion sinks (Electron-Ion Recombination - EIR; Molecular Activated Recombination - MAR) $<0$ integrated over the lower/upper inner/outer divertor legs. The black horizontal bars indicate the ion target flux arriving to the four respective targets. In b) the ion target flux is compared to the ion target flux minus the difference in MAR obtained through post-processing. d) Schematic illustration of the post-processing technique. }
\label{fig:STEP_MAR_MAD_postproc}
\end{figure}

The impact of post-processing on particle (i.e., ion) balance and the $D\alpha$ emission profile on this simulation is shown in figure \ref{fig:STEP_MAR_MAD_postproc}. 

The presented analysis is based on integrating the ion sources and sinks over the entire divertor domain. Since the ionisation region is still attached (outer target) / starts to detach (inner target) in the presented simulation, most MAR and MAD occur outside of the killer flux tube, although their contribution at the killer flux tube is non-negligible and increases in more deeply detached simulations. Although both $D_2^+$ and $D^-$ are post-processed, the $D_2^+$ channel drives the dominant MAR and MAD, with $D^-$ contributing less than 25 \% of the total MAR and MAD.

MAR estimates became similar to, and often larger than EIR, after post-processing, despite electron densities $\approx 10^{21} m^{-3}$ being obtained, an example of which is shown in figure \ref{fig:STEP_MAR_MAD_postproc}b. In this example, the post-processed MAR ion sink, integrated over the inner divertor legs, exceeds the inner target ion fluxes. The relative strength of MAR, compared to EIR, at such high electron densities is remarkable. The relative impact of MAR would probably be reduced at high electron densities, since EIR $\propto n_e^3$ in these electron density regimes. However, the tight baffling results in elevated molecular densities that significantly contribute to MAR over a large spatial region. Furthermore, MAR ion sinks can become significant even if EIR is not present, since the temperature at which MAR can occur ($T < 2.5$ eV) is significantly higher than that of EIR ($T < 1.5$ eV). The relative importance of MAR is consistent with the results of the MAST-U Super-X divertor in figure \ref{fig:PartBalSXD}.

In addition to MAR, post-processing (figure \ref{fig:STEP_MAR_MAD_postproc}) also greatly boosts MAD, boosting the total generation of neutral atoms from dissociation, integrated over the inner / outer divertor legs, by a factor $\times 8-9$ /  $\times 2-2.5$, respectively. This will substantially increase the neutral atom density, which may impact a range of different aspects, including 
\begin{itemize}
    \item Divertor neutral pressure \cite{Verhaegh2023b}.
    \item Crosstalk between the inner and outer divertor \cite{Kukushkin2016}.
    \item Divertor pumping and neutral atom balance \cite{Kukushkin2016}.
    \item Neutral target heat loads, as well as heat transport from charge exchange of hot neutrals. If the additional neutrals travel to a hot part of the divertor leg and undergo charge exchange, they may increase the neutral target heat loads.
    \item Changes in fuelling efficiency. In \cite{Verhaegh2023b}, the same upstream density was reached at a 12 \% lower fuelling rate in deeply detached self-consistent TCV SOLPS-ITER simulations with modified rates, compared to the default rate setup \cite{Verhaegh2023b}.
\end{itemize}

MAR and MAD also lead to non-negligible electron cooling. In the example presented in figure \ref{fig:STEP_MAR_MAD_postproc}, these processes result in $\sim 5$ MW of additional hydrogenic atomic radiative power losses, as well as $\sim 6$ MW of additional electron cooling from the dissociation associated with MAR and MAD; accounting for $\sim 14 \%$ of the $P_{SOL}$. Such estimates are consistent with the results of the MAST-U Super-X divertor in figure \ref{fig:PowerLossSXD}.

In conclusion, post-processing tightly baffled reactor SOLPS-ITER simulations suggest that plasma-molecular chemistry can have a significant impact even on a reactor scale. The increase in ion sinks during detachment as a result of the additional MAR and the increase in power dissipation as a result of the additional MAD are expected to have a significant impact on detachment.

\subsubsection{Impact of plasma-molecular chemistry on diagnostic analysis and development with implications for reactors}

In addition to impacting the divertor state, the additional $D_2^+$ resulting from post-processing significantly impacts hydrogen emission. In figure \ref{fig:STEP_MAR_MAD_postproc}c,  the $D\alpha$ emission increases by a factor $\times \sim 2$ after post-processing. This additional $D\alpha$ emission associated with MAR and MAD can be significant in the far-SOL, broadening the entire hydrogen emission region. 

The strong increase of the $D\alpha$ emission near the X-point, combined with light reflecting off the metal surfaces, can have significant impacts on diagnostic development for reactors. For example, there is concern that the emission of $D\alpha$ results in additional stray light for diagnostics \cite{Kukushkin2016a}, which would become more problematic when plasma-molecular interactions are considered. If hydrogen emission is used to analyse divertor processes in a reactor or for exhaust control, the strong increase in the $D\alpha$ emission due to plasma-molecular interactions, as well as the change in the spatial profile of $D\alpha$ emission, requires the use of analysis tools and control sensor strategies that can include plasma-molecular interactions.

\subsubsection{The interplay between photon opacity and plasma-molecular interactions}
\label{ch:photon_opacity}

In addition to plasma-molecular interactions contributing to atomic hydrogen emission, also photon opacity can change the hydrogen emission significantly and boost $D\alpha$ emission. Photon opacity, which is not taken into account in these simulations \cite{Osawa2023,Hudoba2023}, will be significant under these reactor conditions given the high neutral atom densities obtained and the size of the divertor ($\Delta L \times n_{D} \gg 10^{18} m^{-2}$). The neutral atom density would be further enhanced through plasma-molecular interactions by the additional hydrogen atoms generated through MAD and MAR.

The interplay between plasma-molecular interactions and photon opacity can significantly complicate the diagnosis and interpretation of hydrogen emission signals, as it would result in a spatial separation between the emission and opacity regions of $Ly\alpha$ and $Ly\beta$ \cite{Verhaegh2021a}. The impact of this on diagnostic development, analysis, and real-time sensor capabilities that rely on atomic hydrogenic emission requires further study.

As photon opacity effectively prolongs the time atoms are excited through cycles of (re-)absorption and (re-)emission, and since excited atoms have a lower ionisation energy, re-ionisation events are increased if opacity is high, increasing (decreasing) effective ionisation (recombination) rates and lowering the energy cost of ionisation \cite{Pshenov2023,Lomanowski2019}. Through this mechanism, photon opacity can change the balance between MAR, MAD, and Molecular Activated Ionisation (MAI), as well as the ratio between MAR and EIR. Further collisional-radiative modelling studies are required to understand the impact of photon opacity on MAR, MAD and MAI.


\subsubsection{Practical implications of our reactor scale findings}

From a zero-order perspective, increased MAR and MAD is expected to result in more deeply detached conditions as it results in additional power/particle losses, which are expected to impact detachment significantly. 

However, the impact of this on the plasma solution can result in more complex higher-order effects, such as changes to the electron density profile that can then impact the impurity radiative power losses. Such effects complicate making predictions on how the additional MAR and MAD impact the divertor solution without new, fully self-consistent, simulations. Previous work showed that the impact of molecular charge exchange on momentum balance is complex, having knock-on effects on momentum losses driven through plasma-atom and molecule collisions by increasing/decreasing the local neutral atom/molecular densities because of the additional dissociation driven by MAD.

\begin{itemize}
\item Improved understanding of plasma-molecular interactions, including MAR and MAD, is crucial for reactor design. It highlights the need for self-consistent simulations that account for these interactions.
\item The significant increase in hydrogenic atomic emission, according to post-processing, underscores the importance of refining diagnostic tools and analysis strategies that use hydrogen emission for plasma-molecular interactions. Additionally, the potential impact of plasma-molecular interactions must be considered in diagnostic design.
\item The discussion of uncertainties in the molecular charge exchange rates emphasises the need for more precise rate calculations through experimental, collisional-radiative modelling, and exhaust modelling studies. 
\item The interplay between photon opacity and interactions such as MAR and MAD arising from molecular charge exchange is complex and can impact diagnostic measurements/analysis as well as exhaust physics, warranting further study.
\end{itemize}

\subsection{Relevance of these results and implications for reactors}

This work shows, for the first time, that MAR can be significant in higher power, non-ohmic, conditions. Higher power results on MAST-U (up to $P_{NBI} = 2.5$ MW) highlighted qualitatively similar findings as previous ohmic results \cite{Verhaegh2023,Wijkamp2023,Verhaegh2023b,Moulton2023}, suggesting that the high impact of plasma-molecular interactions observed previously are not only due to the low power conditions. Additional investigation showed that $D^-$ may also be non-negligible under beam-heated MAST-U conditions, although the MAR driven by $D_2^+$ remained dominant.

Although our experimental results show promising benefits of the Super-X divertor, higher-power and H-mode experiments, featuring elevated electron and neutral densities, are essential to understand the behaviour of MAR and MAD at higher power, as well as the physics of alternative divertor configurations. Such studies will be a focal point of MAST-U research \cite{Harrison2023}, with an emphasis on exploring scaling behaviour for reactor-class devices. Higher powers can impact the relevance of plasma-molecular interactions, and extensive research is required to comprehensively assess how alternative divertor configurations scale in the context of future fusion reactors. Further investigations on plasma-molecular effects in other divertor configurations in MAST-U are planned \cite{Verhaegh2023d}.

Nevertheless, the relevance of plasma-molecular chemistry in higher power conditions is supported by post-processing reactor-scale simulations with improved molecular charge exchange rates. This resulted in significant modifications to the divertor state as well as synthetic diagnostic results, requiring accounting for plasma-molecular effects in the development of analysis techniques, real-time control sensing, and diagnostic design.

\section{Conclusions}
\label{ch:conclusion} 

The results of this study offer valuable insights into the intricate role that plasma-neutral interactions play in power and particle exhaust within fusion devices. Research conducted on the MAST-U Super-X divertor, under NBI-heated conditions (up to 2.5 MW of NBI power), has revealed findings that agree with previous observations of the importance of molecular processes in the divertor for ohmic conditions. Although some distinctions, such as higher divertor electron temperatures ($\sim 0.2$ eV for NBI-heated conditions as opposed to $<0.2$ eV for ohmic conditions), have been observed, the overall detachment behaviour seems to scale to higher power conditions on MAST-U. Remarkably, the ionisation source in the divertor chamber demonstrates robustness in response to changes in steady-state values of core density and heating, shifting only marginally when either the core electron density or the absorbed beam power are nearly doubled. This underscores the robustness of the detachment behaviour in the Super-X divertor configuration.

The study has brought to light the pivotal roles of Molecular Activated Recombination (MAR) and Molecular Activated Dissociation (MAD) in power and particle exhaust, showing their continued significance even in NBI-heated scenarios. This suggests that plasma-molecular interactions extend their influence beyond low-power regimes, highlighting their relevance for future fusion reactors. Further analysis has revealed the potential contributions of $D^-$. This is consistent with the effective rates computed using ab initio electron attachment cross-sections for deuterium ($e^- + D_2 \rightarrow D_2^- \rightarrow D^- + D$). This additional complexity has implications for the overall behaviour of fusion plasmas and underscores the need for further exploration.

To assess the implications of these interactions in a reactor context, reactor-scale simulations of a tightly baffled divertor that exhibits alternative divertor geometries have been post-processed with newly calculated effective molecular charge exchange ($D_2 + D^+ \rightarrow D_2^+ + D$) and dissociative attachment ($e^- + D_2 \rightarrow D_2^- \rightarrow D^- + D$) rates. The results from this non-self-consistent scoping study demonstrated the importance of MAR, compared to Electron-Ion Recombination (EIR), even in the presence of high electron densities (around $10^{21} m^{-3}$). Additionally, MAD was shown to significantly enhance the total dissociation rate by up to a factor of eight. The power losses associated with MAR and MAD were significant, accounting for 10-20\% of the power entering the separatrix. These interactions can result in a doubling to tripling of the $D\alpha$ emission, which can have significant impacts on diagnostic analysis and development.

This study advances our understanding of plasma-molecular interactions and their impact on power exhaust. These insights are invaluable for informing the design of future fusion reactors and the development of diagnostic tools. The complex and multifaceted nature of these interactions underscores the need for continued research, modelling, and experimental investigations to fully understand their implications in a range of divertor plasma scenarios. Self-consistent simulations that account for these interactions will be crucial to achieving more accurate predictions in reactor design. 

\section{Acknowledgements}

This work has received support from EPSRC Grants EP/T012250/1 and EP/N023846/1. This work has been carried out within the framework of the EUROfusion Consortium, partially funded by the European Union via the Euratom Research and Training Programme (Grant Agreement No 101052200 — EUROfusion). The Swiss contribution to this work has been funded by the Swiss State Secretariat for Education, Research and Innovation (SERI). The work by S. Kobussen was supported by FuseNet. The results are obtained with the help of the EIRENE package (see www.eirene.de) including the related code, data and tools \cite{Reiter2005}. Views and opinions expressed are however those of the author(s) only and do not necessarily reflect those of the European Union, the European Commission or SERI. Neither the European Union nor the European Commission nor SERI can be held responsible for them. 

\section{Data availability statement}

The data that support these studies are openly available at DOIs:

\begin{enumerate}

\item Miscellaneous (standard) information from MAST-U for discharges \# 46860, 46705, 47958, 48330, including Langmuir probe derived particle fluxes. DOI: \url{https://doi.org/10.14468/h7np-q868}. Figures \ref{fig:ProfsSXD},\ref{fig:PartBalSXD},\ref{fig:D2Fulcher_XPT},\ref{fig:PowerLossSXD},\ref{fig:PowerTransientSXD},\ref{fig:MAR_Dm}.

\item Default Bayesian spectroscopy (BaSPMI) analysis \# 46860. DOI: \url{https://doi.org/10.14468/bfcj-j942}. Figures \ref{fig:ProfsSXD},\ref{fig:PartBalSXD},\ref{fig:PowerLossSXD}.

\item Imaging Bolometer (IRVB) analysis \# 47958. DOI: \url{https://doi.org/10.14468/k38c-4687}. Figures \ref{fig:D2Fulcher_XPT},\ref{fig:PowerLossSXD}.

\item Combined inversions Multi-Wavelength Imaging and X-Point Imaging of $D_2$ Fulcher emission \# 48330. DOI: \url{https://doi.org/10.14468/q3q5-d724}. Figure \ref{fig:D2Fulcher_XPT}.

\item Default Bayesian spectroscopy (BaSPMI) analysis \# 46705. DOI: \url{https://doi.org/10.14468/fmbr-d4272}. Figure \ref{fig:PowerTransientSXD}.

\item Custom Bayesian spectroscopy (BaSPMI) analysis \# 46705, including $D^-$ effects. DOI: \url{https://doi.org/10.14468/rz52-cm67}. Figure \ref{fig:MAR_Dm}.

\item Modified rate calculations using CRUMPET collisional-radiative modelling. DOI: \url{https://doi.org/10.14468/3ybq-jh97}. Figures \ref{fig:DmD2p_MAR},\ref{fig:fD2v}.

\item Post-processing of STEP simulation for molecular effects. DOI: \url{https://doi.org/10.14468/x4p5-b104}. Figure \ref{fig:STEP_MAR_MAD_postproc}.

\end{enumerate}

To obtain further information on the data and models underlying this paper please contact publicationsmanager@ukaea.uk.




\section{References}
\bibliographystyle{iopart-num}
\bibliography{all_bib.bib}

\appendix

\section{Rate calculations using collisional-radiative modelling of $D_2 (\nu)$ as a weighting function}
\label{ch:colrad}

A vibrationally resolved CRM is used, analogously to the Eirene approach, to compute the vibrational distribution, assuming that the molecules are in both the rotational and electronic ground states. This is used as a weighting function for vibrationally resolved rates to calculate effective rates ($<\sigma v>_{eff} (T) = \Sigma_\nu <\sigma v>_{\nu} (T) f_\nu(T)$). The reactions used in this CRM are specified in table \ref{tab:Reacs} and the CRUMPET tool \cite{Holm2022} is used to implement the CRM.

\begin{table}[h!]
\begin{tabular}{lll}
Description                              & Reaction                                                                                                                        & Data                                                  \\
Vibrational excitation  & $e^- + H_2 (\nu) \rightarrow e^- + H_2 (\nu \pm 1)$                                                                             & Eirene / 'H2VIBR'  \cite{Reiter2005} \\
$H_2$ ionisation                         & $e^- + H_2 (\nu) \rightarrow 2 e^- + H_2^+$                                                                                     & Eirene / 'H2VIBR'  \cite{Reiter2005} \\
$H_2$ dissociation                       & $e^- + H_2 (\nu) \rightarrow e^- + H + H$                                                                                        & Eirene / 'H2VIBR' \cite{Reiter2005}  \\
Molecular charge exchange                & $D^+ + D_2 (\nu) \rightarrow D_2^+ +D$                                                                                         & Ichihara \cite{Ichihara2000}         \\
Dissociative attachment                & $e^- + D_2 (\nu) \rightarrow D^- + D$                                                                                         & Laporta \cite{Laporta2021}
\end{tabular}
\caption{Table showing reactions used for collisional-radiative modelling. Table adopted from \cite{Verhaegh2023b}.}
\label{tab:Reacs}
\end{table}

It is assumed that the hydrogen rates for vibrational excitation, $H_2$ ionisation, and dissociation, all three for which the default Eirene data are used, are identical to the deuterium rates. Data for deuterium is used for dissociative attachment \cite{Laporta2021}. Since the molecular charge exchange rates are similar between the different isotopes \emph{at the same collision velocity and the same energy level of the vibrational state} \cite{Ichihara2000,Reiter2018}, they are remapped from hydrogen ($E_{H_2, \nu}$) to deuterium ($E_{D_2, \nu}$). This means that the resonance transition for $H_2$ at $\nu\geq 4$ (where the molecular charge exchange rates are strongly increased) is shifted to $\nu \geq 6$ for $D_2$. Due to the higher isotope mass of deuterium, the lower relative velocity between $D^+$ and $D_2$ (assumed to be static at 0.1 eV) at the same ion temperature has been taken into account by employing ion isotope mass rescaling to the ion temperature ($<\sigma v>_{D_2} (E_{D_2, \nu}, T) = <\sigma v>_{H_2} (E_{D_2, \nu}, T/2)$). Eirene, instead, applies this rescaling to the effective rates, hence inadvertently rescaling the full $f_\nu(T)$ dependency (including interactions that do not depend on ion temperature); see \cite{Verhaegh2023a,Verhaegh2023b}.

The CRM assumes that the electron temperature is equal to the ion temperature $T=T_e=T_i$ and that the electron density is equal to the density $H^+$. In this collisional-radiative model calculation, $D_2(\nu = 0)$, $D$, $e^-$, $D^+$, $D_2^+$ and $D^-$ have been set up as reservoir species. The time-dependent CRM is solved numerically until a quasi-steady state is obtained from which $f_\nu$ is adopted (for $\nu=1$ to $\nu=14$). 

\begin{figure}
\centering
\includegraphics[width=\linewidth]{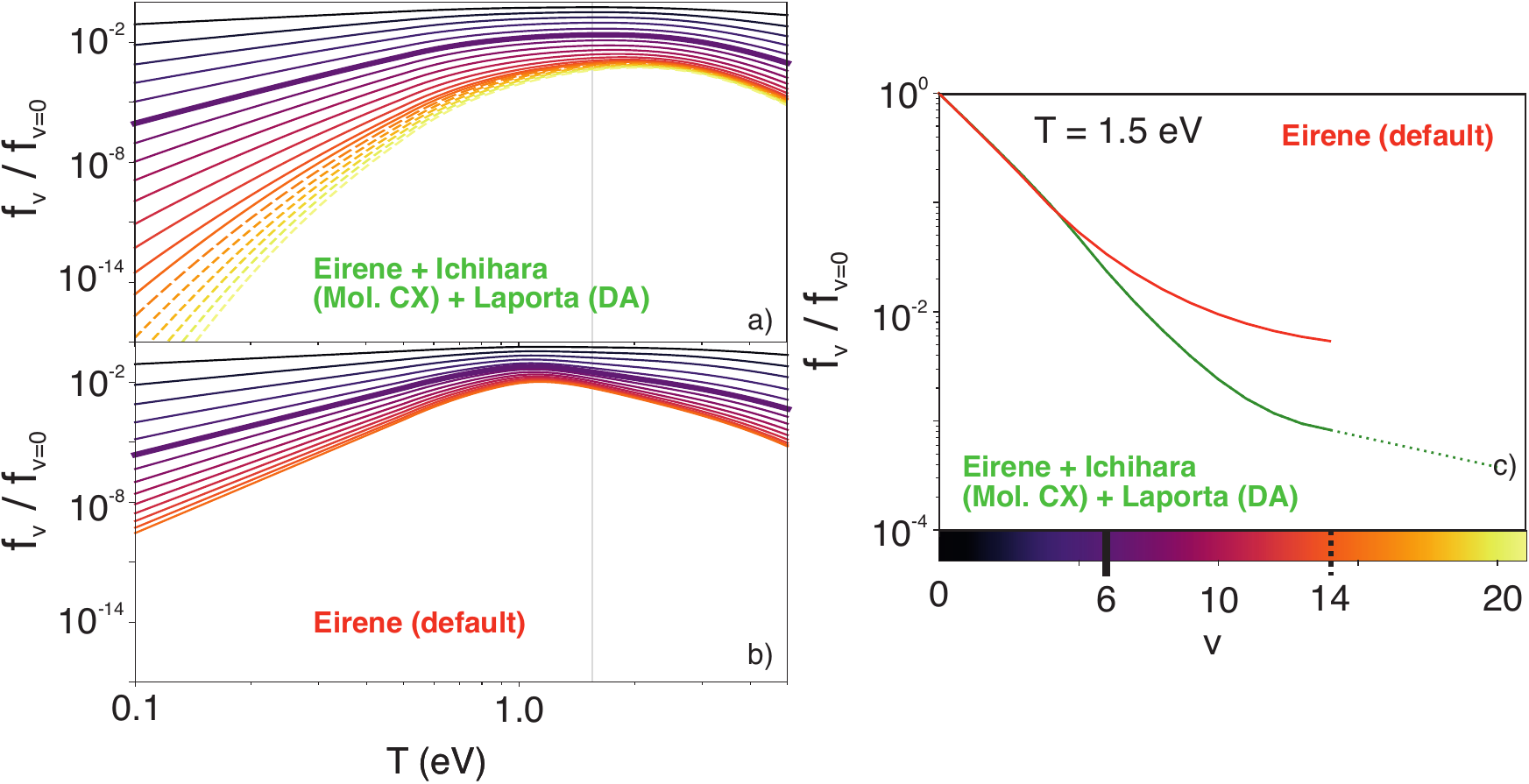}
\caption{Result from the vibrational resolved CRM. a) $f_\nu (T)$ for the CRM indicated in table \ref{tab:Reacs} for $\nu=1$ up to $\nu=20$, for which $\nu>14$ is extrapolated (dotted lines). b) $f_\nu (T)$ for the Eirene (default) CRM (without dissociative attachment and with molecular charge exchange changed to the Eirene / 'H2VIBR' \cite{Reiter2005}). c) $f_\nu$ vs $\nu$ at $T=1.5 eV$, including extrapolation of the vibrational distribution. The colourbar at the x-axis is related to the vibrational levels in figures a, b. $f_{\nu=6}$ is highlighted in figure a, b as this corresponds to the resonance level for molecular charge exchange.}
\label{fig:fD2v}
\end{figure}

Figure \ref{fig:fD2v}a shows the solution of $f_\nu (T)$, relative to $D_2 (\nu=0)$ from our CRM, while \ref{fig:fD2v}b shows the Eirene solution (where dissociative attachment is neglected and the Eirene molecular charge exchange rate is used). This shows, indeed, that dissociative attachment and increased molecular charge exchange results in additional depletion of higher vibrationally excited levels, which is accounted for self-consistently in our effective rate calculations. The vibrational distribution obtained from our CRM (up to $\nu=14$) is extrapolated to higher vibrational states (up to $\nu=20$), as the dissociative attachment cross-sections increase strongly with increased vibrational number. This is shown in figure \ref{fig:fD2v}c, where the vibrational distribution is shown on a logarithmic scale including the log-linear extrapolation to higher vibrational states.







\end{document}